\documentclass[twocolumn]{aastex631}
\usepackage{dcolumn}
\usepackage{xcolor}
\usepackage{amsthm,amsmath,amssymb}
\usepackage{mathrsfs}

\newcolumntype{C}[1]{>{\centering\arraybackslash}p{#1}} % 居中对齐
\newcolumntype{L}[1]{>{\raggedright\arraybackslash}p{#1}} % 靠左对齐

\newcommand{\lya}{Ly$\alpha~\lambda$1216}
\newcommand{\Lya}{Ly$\alpha$}
\newcommand{\nv}{N~{\sc v$~\lambda$}1240}
\newcommand{\Nv}{N~{\sc v}}
\newcommand{\lyanv}{Ly$\alpha$~$+$~N~{\sc v}}
\newcommand{\oi}{O~{\sc i$~\lambda$}1304}
\newcommand{\Oi}{O~{\sc i}}
\newcommand{\civ}{C~{\sc iv$~\lambda$}1549}
\newcommand{\Civ}{C~{\sc iv}}
\newcommand{\heii}{He~{\sc ii$~\lambda$}1640}
\newcommand{\Heii}{He~{\sc ii}}

\newcommand{\ciii}{C~{\sc iii]$~\lambda$}1909}
\newcommand{\Ciii}{C~{\sc iii]}}
\newcommand{\mgii}{Mg~{\sc ii$~\lambda$}2799}
\newcommand{\Mgii}{Mg~{\sc ii}}

\newcommand{\Hb}{H$\beta$}
\newcommand{\oiii}{[O~{\sc iii}]~$\lambda$5007}
\newcommand{\Oiii}{[O~{\sc iii}]}
\newcommand{\ha}{H$\alpha~\lambda$6563}

\newcommand{\feii}{Fe~{\sc ii}}

\newcommand{\siiv}{Si~{\sc iv}}

\newcommand{\ledd}{$\lambda_{\rm Edd}$}
\newcommand{\mytt}[1]{\texttt{#1}}

\graphicspath{{./}{figures/}}

\begin{document}

% \uppercase\expandafter{\romannumeral4}

\title{Questing a Coherent Definition of Weak-line Quasars and its Physical Implications}

\author[0009-0005-2621-2479]{Xiaoqiang Cheng}
\affil{Department of Astronomy, Xiamen University, Xiamen, Fujian 361005, China}
\author[0000-0001-7349-4695]{Jianfeng Wu}
\affil{Department of Astronomy, Xiamen University, Xiamen, Fujian 361005, China}
\author[0000-0003-4202-1232]{Qiaoya Wu}
\affil{Department of Astronomy, University of Illinois at Urbana-Champaign, Urbana, IL 61801, USA}

\correspondingauthor{Jianfeng Wu}
\email{wujianfeng@xmu.edu.cn}

\begin{abstract}
Weak-line quasars (WLQs) represent a subset of type 1 quasars distinguished by remarkably weak high-ionization broad emission lines, yet exhibiting normal optical/UV continua. This study establishes a physically motivated definition of WLQs using 371,091 quasars from the Sloan Digital Sky Survey Data Release 16 catalog. By analyzing outliers in three key relations, the \mbox{$L_{1350\rm \textup\AA}-$\rm \Civ} blueshift relation, the Baldwin effect, and the \mbox{$\log L_{2500\textup\AA}-\alpha_{\rm ox}$} relation, we identify two critical thresholds in C~{\sc iv} equivalent width (EW): $8.9\pm0.2$~\AA\ and $19.3\pm0.3$~\AA. Quasars with EW(\Civ)~\textless~$8.9\pm0.2$~\AA\ are defined as WLQs, exhibiting enhanced C~{\sc iv} blueshifts, significant deviations from the Baldwin effect, and a high fraction of X-ray weak objects (nearly half of this population). Quasars with EW(\Civ)~\textgreater~$19.3\pm0.3$~\AA\ show normal quasar properties, while objects with intermediate \Civ\ equivalent width ($8.9-19.3$~\AA) are defined as ``bridge quasars'', showing transitional behaviors. Systematic analysis of emission-line attenuation of WLQs reveals a clear positive correlation between the attenuation factor and ionization energy, with high-ionization lines (e.g., \Heii, \Civ) suppressed by factors of $\sim3$--4$\sigma$ compared to low-ionization lines (e.g., \Mgii, \Oi). This ionization-stratified attenuation supports the shielding gas model, where geometrically thick inner accretion disk obscures high-energy photons, suppressing high-ionization line emission, while the low-ionization lines remain less affected. 
%Additionally, WLQs show marginally higher Eddington ratios (\ledd $\sim0.15$) than normal quasars (\ledd $\sim0.12$), though potential biases in single-epoch black hole mass estimates suggest true accretion rates may be higher. 
% Our results unify the multiwavelength properties of WLQs and provide a robust framework for studying super-Eddington accretion and early SMBH growth.
Based on this accretion-disk geometry, we argue that WLQs and normal quasars defined in our framework generally correspond to the slim disk and standard thin disk regimes, respectively, while the bridge quasars represent a transitional phase betweeen the two states.
This work establishes a unified observational criterion for WLQs and discusses their physical implications, highlighting the role of accretion-driven shielding gas in shaping their unique spectral features.

\end{abstract}


\section{Introduction} \label{sec:intro}
Strong broad emission lines are one of the prominent features of type 1 quasar spectra in the optical/UV bands. However, a subset of radio-quiet quasars with remarkably weak broad emission lines but blue continuum, termed weak-line quasars (WLQs), has been identified in recent years. 
The first identification of a WLQ at high redshift (SDSS~J1530$-$00; $z=4.62$) was reported by \cite{1999ApJ...526L..57F}, which has a featureless optical spectrum from the Sloan Digital Sky Survey (SDSS; \citealt{2000AJ....120.1579Y}). The low-redshift quasar PG~1407$+$265, exhibits notably weak emission lines, such as Ly$\alpha~\lambda$1216 and \civ, while displaying normal \mgii\ and \ha\ emissions \citep{1995ApJ...450..585M}. More individual WLQs have since been identified and/or extensively studied based on their optical/UV spectra (e.g., SDSS~J1032$+$0300, \citealt{2001AJ....122..503A};  SDSS~J0040$-$0915, \citealt{2003AJ....126.2579S}; 2QZ~J2154$-$3056, \citealt{2004MNRAS.352..903L}; SDSS~J1335$+$3533, \citealt{2006AJ....131.1203F}; PHL~1811, \citealt{2007ApJ...663..103L,2007ApJS..173....1L}). 

Systematic searches of WLQs were carried out by \citet{2009ApJ...699..782D} and \citet{2010AJ....139..390P} for high-redshift ($z>3$) and low-redshift ($z < 2.2$) objects, respectively, both of which were based on the SDSS spectroscopic database. The physical nature of WLQs have been investigated based on the multiwavelength data of these samples. The BL Lac-like object scenario were ruled out based on their radio weakness \citep{2010ApJ...721..562P} and multi-band spectral energy distributions (SEDs; \citealt{2011ApJ...743..163L}). X-ray properties of WLQs have been extensively studied (e.g., \citealt{2009ApJ...696..580S,2011ApJ...736...28W,2012ApJ...747...10W,2015ApJ...805..122L,2018MNRAS.480.5184N,2020ApJ...889L..37N,2022MNRAS.511.5251N}). It has been demonstrated that WLQ samples are characterized by the exceptionally high fraction ($\sim50\%$) of X-ray weak objects\footnote{Following \cite{1979ApJ...234L...9T}, we define the X-ray-to-optical power-law slope parameter $\alpha_{\rm ox}$, calculated as \mbox{$\alpha_{\rm ox}=0.3838\log(L_{2~{\rm keV}}/L_{2500~{\textup{\AA}}})$}, to quantify the ratio between the X-ray luminosity in rest-frame 2~keV and optical/UV luminosity at rest-frame 2500~\AA ~of AGNs. And we adopt $\Delta\alpha_{\rm ox}$ parameter, calculated by $\Delta\alpha_{\rm ox}=\alpha_{\rm ox}-\alpha_{\rm ox,exp}$, to quantify the degree to which $\alpha_{\rm ox}$ deviates from the expectation based on the $\alpha_{\rm ox}$-$L_{2500~{\textup{\AA}}}$ relation in \citet{2007ApJ...665.1004J}. We define a quasar with $\Delta\alpha_{\rm ox} \textless -0.2$ as X-ray weak following prior WLQ studies (e.g., \citealt{2015ApJ...805..122L, 2018MNRAS.480.5184N}).} \citep{2012ApJ...747...10W,2015ApJ...805..122L}, in contrast of less than $\sim10\%$ for the normal quasar population \citep{2008ApJ...685..773G,2020ApJ...900..141P}. The X-ray weak WLQs have shown hard X-ray spectra, indicating heavy X-ray absorption. The absorbing materials need to reside near the X-ray emitting region, since the optical/UV spectra do not show signature of obscuration \citep{2011ApJ...736...28W}. On the other hand, X-ray normal WLQs often have a steep X-ray power law ($\Gamma > 2$), suggesting they may have high Eddington ratio ($\lambda_{\rm Edd} = L/L_{\rm Edd}$; e.g., \citealt{2008ApJ...682...81S,2018ApJ...865...92M}).  

Near-infrared spectroscopy of high-redshift WLQs, combined with optical/UV spectra, provides a more complete picture of the emission-line properties of WLQs. The distinctive characters of WLQs are their exceptionally weak high-ionization lines (HILs; e.g., \Civ, \siiv, and \Heii), while their low-ionization lines (LILs; e.g., \Mgii) have similar strength to typical quasars (e.g., \citealt{2015ApJ...805..123P}). Furthermore, HILs, especially the \Civ\ line of WLQs often show significant blueshifts (typically exceeding 1000~km~s$^{-1}$; e.g., \citealt{2011ApJ...736...28W,2015ApJ...805..122L}), indicating  strong outflows from the quasars. The weak line strength and strong blueshift of the \Civ\ emission are considered to be signatures of high Eddington ratios \citep{2011AJ....141..167R,2022ApJ...931..154R,2023ApJ...950...97H}, which are also connected to the X-ray weakness of quasars. 

Several scenarios have been proposed to explain the line weakness of WLQs. One is the ``anemic" broad line regions (BLRs), i.e., the deficit of line-emitting gas results in the weak lines (e.g., \citealt{2010ApJ...722L.152S}). In this case, WLQs could be manifestations of quasars in their early formation stages that their BLRs have not been fully developed \citep{2010MNRAS.404.2028H,2011ApJ...728L..44L,2020ApJ...903...34A}. However, this scenario predicts that all broad emission lines should be weak which is at odds with the observed fact that WLQs’ LILs exhibit strengths comparable to those of typical quasars. Alternatively, the BLR of WLQs could be receiving a much softer ionizing continuum, which leads to unusually weak HILs, while the LILs are less affected. \cite{2011MNRAS.417..681L} proposed that an inefficient cold accretion disk caused by a very high black hole mass ($\textgreater 3\times 10^{9}\ M_{\odot}$ for a non-rotating black hole) could generate a non-ionizing continuum. However, WLQs that have black hole mass measured do not host such massive black holes (e.g., \citealt{2015ApJ...805..123P}) as \citet{2011MNRAS.417..681L} predicted. Furthermore, these measurements from single-epoch spectroscopy could even overestimate the black hole mass for the high accretion rate WLQs \citep{2018ApJ...856....6D,2022MNRAS.515..491M,2023ApJ...950...97H}. 

On account of the unusual X-ray properties of WLQs, a ``shielding gas" model was proposed to correlate the multiwavelength observations of WLQs into one coherent picture, which can naturally unify the X-ray weak and X-ray normal sub-populations \citep{2011ApJ...736...28W,2012ApJ...747...10W,2015ApJ...805..122L,2018MNRAS.480.5184N}. Due to their high accretion rates, WLQs have geometrically thick inner accretion disks (as in the ``slim disk'' model; \citealt{1988ApJ...332..646A,2019Univ....5..131C}) and the associated outflows \citep{2014ApJ...797...65W,2014ApJ...796..106J,2019ApJ...880...67J}. The thick inner disk and outflows act as ``shielding gas" with a high covering factor of the broad line region (BLR), which prevents the high-energy X-ray photons from reaching the BLR. If the quasar is viewed with an ``edge-on" perspective, i.e., the line of sights going through the thick disk and outflows, the observed X-ray emission is heavily absorbed, while for the ``face-on" case, the X-ray emitting region can be directly viewed and thus the quasar appears as X-ray normal. The BLR always receives a soft ionizing continuum, resulting in weak HILs regardless of orientation, while the LILs are less affected (see Fig.~1 of \citealt{2018MNRAS.480.5184N} for the illustration of the shielding gas model). It is worth noting that although the shielding gas model can adequately explain the correlations of WLQ multiwavelength properties, the entire population of WLQs may contain objects with different physical mechanisms. 

It has been found that the fraction of WLQs among quasars increases significantly with redshifts ($\sim15\%$ at $z>6$, compared to a few percent at lower redshifts; \citealt{2014AJ....148...14B,2016ApJS..227...11B}), which indicates that the supermassive black holes in the early universe are more likely to growth with high accretion rates. This would help alleviate the problem of how supermassive black holes (SMBHs) at high redshifts are formed within very short timescales. 

While WLQs have been proved to be a valuable tool to investigate the quasar geometry, super-Eddington accretion, and the cosmological evolution of SMBHs, one foremost question remains in this area: the lack of a universally adopted definition of WLQs. 
A qualitative description is that WLQs are a class of type 1 quasars having remarkably weak HILs but slightly weak or normal LILs, with the continuum profile similar to that of normal type 1 quasars. Clearly, such a definition cannot effectively guide the WLQ sample selections and subsequent studies. This is partly because of the variations of emission line coverage by the ground-based optical spectroscopy for quasars at different redshifts. \citet{2009ApJ...699..782D} selected WLQs at $z>3$ based on the asymmetric equivalent width (EW) distribution of the \Lya\ + \Nv\ emission line. They defined WLQs as quasars with EW(\Lya\ + \Nv) $< 15.4$~\AA, representing the $3\sigma$ skew tail towards low EW values. Similarly, the distribution of EW(\Civ) also shows asymmetric profile, with EW(\Civ) $< 10$~\AA\ representing the $3\sigma$ weak-line tail. However, as mentioned in \citet{2009ApJ...699..782D} and further demonstrated in \citet{2012ApJ...747...10W}, these two selection criteria based on EW(\Lya\ + \Nv) and EW(\Civ), respectively, are not compatible with each other: a significant portion of WLQs selected based on one criterion do not satisfy the other (see the detailed discussions in Section 3.2 of \citealt{2012ApJ...747...10W}). Studies on low-redshift ($z<2.2$) WLQs are mainly based on the samples selected by \citet{2010AJ....139..390P} that require all emission lines having EW $<5$~\AA, which could involve the \Civ, \Mgii, and/or the Balmer lines. \citet{2022ApJ...929...78P} investigated the \Lya\ + \Nv\ emission strength for six low-redshift WLQs in \citet{2010AJ....139..390P} and found that only two of them satisfy the criterion of EW(\Lya\ + \Nv) $< 15.4$~\AA. The discrepancy in the WLQ selection criteria could undermine the robustness of the comparative studies on WLQs and normal type 1 quasars.
%\cite{2018MNRAS.480.5184N} divided their representative sample into two equal parts, extreme WLQs and bridge quasars, with criteria of EW(\Civ) $\textless$ 7 \AA ~for extreme WLQs and 7 \AA ~$\textless$ EW(\Civ) $\textless$ 15 \AA ~for bridge quasars. These simple WLQ criteria are also somewhat arbitrary and dependent on their sample.

In light of the above, the main purpose of this work is to establish a universal definition of WLQs which reflects their physical nature. Meanwhile, we will quantitatively study the emission line attenuation factor of WLQs and normal quasars in the optical/UV band, to further assess which model(s) can better explain the physical nature of WLQs. Our sample is compiled in Section~\ref{sec:sample}, and the emission-line strength distributions of these quasars are presented in Section~\ref{fitting}. We describe our methodology of classifying WLQs and provide the final WLQ definition in Section~\ref{sec:results}. In Section~\ref{sec:other_emission_lines}, we use our newly selected WLQ sample to study their emission line properties and their physical implications. We discuss the future perspectives of this study in Section~\ref{sec:discuss}, and summarize our conclusions in Section~\ref{sec:summary}. Throughout this work, we use J2000 coordinates and a flat $\Lambda$CDM cosmology with $H_{0}=70 \rm ~km ~s^{-1}\ Mpc^{-1}$, $\rm \Omega_{M}=0.3$ and $\rm \Omega_{\Lambda}=0.7$.

\section{Data Compilation} \label{sec:sample}

This study is based on the value-added catalog for the SDSS Data Release 16 (DR16) quasars\footnote{ http://quasar.astro.illinois.edu/paper\_data/DR16Q/} (\citealt{2022ApJS..263...42W}; WS22 hereafter). The WS22 catalog provides the continuum and emission line properties for the largest selection of spectroscopically confirmed broad-line quasars by the SDSS to date, totaling 750,414 (the DR16Q catalog; \citealt{2020ApJS..250....8L}). The quasar spectroscopy is conducted using the BOSS spectrographs on the 2.5-meter Sloan Telescope. Each of the two spectrographs collected data from 500 fibers on a 2k CCD with 24 $\rm\mu$m square pixels, covering a wavelength range from 3600 to 10400 \AA ~with a spectral resolution of $\lambda / \Delta \lambda \approx 2000$. 
%Continuum and emission line properties of these 750414 type 1 quasars are well compiled in WS22 catalog, and we will use these properties for our work directly after data reduction.

\subsection{Sample Selection}\label{sec:general}

The primary objective of this section is to compile a highly reliable and unbiased quasar sample for subsequent analyses by excluding objects that are radio-loud, exhibit strong absorption features, or possess spectra with low signal-to-noise ratios ($S/N$). We first exclude 5 quasars having negative or null values for systemic redshift. 

For radio-loud quasars, we adopt \mbox{$R=f_{\rm \textup5GHz}/f_{2500\textup{\AA}}$} as the radio-loudness parameter (e.g., \citealt{2007ApJ...656..680J}), where $f_{\rm5GHz}$ and \mbox{$f_{2500\textup{\AA}}$} are flux density values at rest-frame 5 GHz and 2500~\AA, respectively. The monochromatic luminosity at rest-frame 2500~\AA\ is provided in WS22 catalog, which is converted to \mbox{$f_{2500\textup{\AA}}$} with corresponding systemic redshift. Then we obtained $f_{\rm5GHz}$ of each source from the flux density at observed-frame 1.4 GHz given by the Faint Images of the Radio Sky at Twenty centimeter survey (FIRST; \citealt{1995ApJ...450..559B}) and the NRAO Very Large Array Sky Survey (NVSS; \citealt{1998AJ....115.1693C}) after applying $K$-correction. 
The DR16Q catalog has already provided the radio flux density from the 2014 December version of the FIRST\footnote{https://sundog.stsci.edu/first/catalogs.html} survey catalog (the \mytt{FIRST\_FLUX} column).  
Assuming a power-law radio continuum of $f_\nu \sim \upsilon^{-0.5}$, we converted the observe frame 1.4 GHz flux density of each matched sources into $f_{\rm5GHz}$ with corresponding systemic redshift. We set $\rm R \textgreater10$ as the threshold for radio-loud quasars and only consider sources with confirmed radio detections, excluding 18,519 objects from our study. 
As a complement, we also matched the DR16Q catalog with NVSS catalog using a $30\arcsec$ radius. Following same calculations, 28,089 radio-loud sources were identified in the NVSS catalog. After removing duplicates, 31,946 radio-loud objects were excluded in total. 

%DLAs are the population of strong absorbers with integrated neutral hydrogen (H~{\sc i}) column density $N_{HI}$ $\geq 2 \times 10^{20} cm^{-2}$ \citep{2005ARA&A..43..861W}, resulting in a broad absorption region ($\Delta v \sim 10^{3} km/s$). 

In the DR16Q catalog, the columns of \mytt{CONF\_DLA} and \mytt{BAL\_PROB} represent the confidence of detection for damped Ly$\alpha$ systems (DLA) and broad absorption line (BAL) quasars, respectively. Following \cite{2020ApJS..250....8L}, we use \mytt{CONF\_DLA $\neq -1$} and \mytt{BAL\_PROB $\geq 0.75$} as the criteria for DLAs and BAL quasars, respectively. Accordingly, we excluded 35,686 DLAs and 99,699 BAL quasars. The number of DLAs in this study is consistent with \cite{2020ApJS..250....8L}, but the number of BAL is slightly less than in \cite{2020ApJS..250....8L}. It should be noted that we only exclude the quasars that cover \lya\ and \civ\ in the SDSS spectra and are confirmed as DLAs and BALs. Finally, we choose sources with high spectral quality based on the \mytt{SNR\_MEDIAN\_ALL} value in the WS22 catalog, which represents the median $S/N$ across all good pixels in a spectrum. We require \mytt{SNR\_MEDIAN\_ALL $\geq 3$} to remove sources with low signal-to-noise ratio, excluding a total of 280,808 sources. Table~\ref{Criterion} lists the number of sources excluded by each criterion, as well as the total number of excluded sources after accounting for duplications. All the subsequent analyses are conducted using the remaining sample of 371,091 objects.

\begin{deluxetable}{C{4cm} C{4cm}}[t!]
	\tabletypesize{\small}
	\tablecaption{Sample Selection Criteria \label{Criterion}}
	\tablehead{
		\colhead{Criterion} & 
		\colhead{Number of removed sources} 
	}
	\startdata
	Redshift $ > 0$ &  5\\  
	Non-radio-loud &  31,946\\
	Non-DLAs &  35,686\\
	Non-BALs & 99,699 \\
	SNR $\geq 3$ & 280,808 \\
	\hline
	Total & 379,323
	\enddata
	%\tablecomments{Note that some sources may meet more than one criterion, so the total number of excluded sources does not equal the sum of the number of excluded sources for each criterion.}
\end{deluxetable}

\subsection{Selection Criteria for Specific Emission Lines}\label{subsec:specific}

This study is largely based on the properties of individual emission line (e.g. Ly$\alpha~\lambda$1216, C~{\sc iv$~\lambda$}1549; see Section~\ref{fitting}), necessitating the screening of quasar spectra containing high-quality emission lines. To this end, we adopted following recommended quality cuts in WS22 for each individual emission line: 
\begin{enumerate}
	\item[$\bullet$] line flux / flux error $> 2$\ ,
	\item[$\bullet$] $38< \log (L_{\rm line}/{\rm erg\,s^{-1}})<48$\ ,
	\item[$\bullet$] $N_{\rm pix, line\ complex} > 0.5 \times N_{\rm max}$\ ,
\end{enumerate}
where $N_{\rm max}$ and $N_{\rm pix, line\ complex}$ are maximum number of available pixels and actual observed pixels in SDSS spectra for a given line complex. The first criterion (i.e., line detection confidence level at $\textgreater 2\sigma$) excludes noisy and potential biased line measurements. Given that the majority of emission lines have luminosities around $\log (L_{\rm line}/{\rm erg\,s^{-1}})\sim 43$, our second criterion in line luminosity range is applied to filter out unphysical values from spectral fitting artifacts; and the third criterion with a pixel number cut is used to remove unconstrained fittings caused by gaps or edge cutoff in the spectra.
% The last two criteria are used to remove unconstrained line due to data gaps in the spectrum.

%Considering that the signal-to-noise ratio will decrease at the edge of SDSS optical coverage, introducing significant scatter, we added another redshift selection criterion. Based on the redshift range of the emission line obtained from the SDSS spectral coverage, we shrink the redshift range by approximately 0.2 to exclude edge regions where the EW measurement is less certain. 

%\section{WLQs as outliers in classical relations} \label{sec:results}

\section{The Equivalent Width Distributions of Broad and Narrow Emission Lines}\label{fitting}

\subsection{High-ionization Broad Lines (HILs)}\label{HILs}

\begin{figure*}[]
	\includegraphics[scale=0.50]{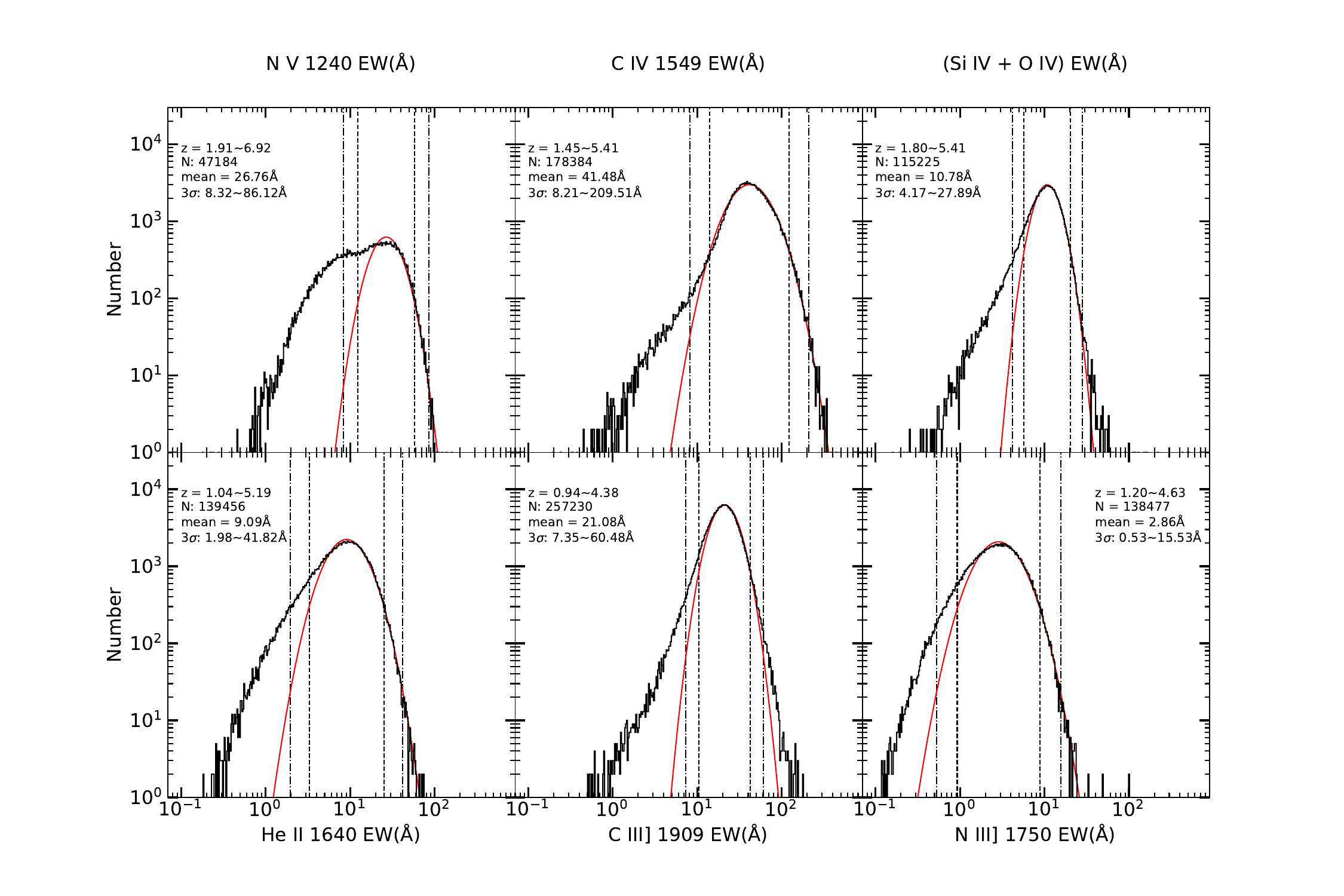}\caption{EW distributions of HILs. In each panel, the black histogram and red lines are the EW distribution and its best-fit lognormal model, respectively. The vertical dash and dash-dotted lines are the $2\sigma$ and $3\sigma$ lines of their lognormal distributions, respectively. In addition, the redshift coverage of each HIL, the number of sources, mean value and the $3\sigma$ range of the best-fit lognormal distribution are depicted in the upper left corner of each panel. As the figure shows, each HIL EW distribution exhibits a prominent tail toward low EW values, which are consistent with the results in \cite{2009ApJ...699..782D} and \cite{2012ApJ...747...10W}.}
	\label{HILs_distributions}
\end{figure*}

Following the approach in \cite{2009ApJ...699..782D} and \cite{2012ApJ...747...10W}, we used a lognormal distribution to fit the EW distribution for each HIL, with all EW values obtained from the WS22 catalog.
The ionization energy values for given ions are retrieved from Table~4 of \citet{2001AJ....122..549V}. Given that \mgii ~is generally viewed as LIL while \ciii\ is generally viewed as HIL, we define the ionization energy of 40~eV as the critical threshold to demarcate HILs from LILs to maintain consistency with established conventions. Due to the different wavelength of each HIL in the rest frame, the number of objects and the covered redshift range for each specific line are also different.\footnote{For each emission line, we shrink the redshift coverage by approximately 0.2 to exclude cases that the line falls within the low-$S/N$ edge regions of SDSS spectra.} 

The EW distributions of HILs are generally asymmetric, exhibiting a tail at the weak end but not at the strong end, which could significantly impacts our fitting results with a single lognormal model. 
%However, the fraction of such tails is generally very small. Our main goal is to fit the part of the distribution where the number of sources accounts for the largest fraction, i.e., its peak and the right side (in logarithmic coordinates).
Therefore, an iterative fitting strategy is employed. 
To mitigate the asymmetric low‑EW tail, we apply an iterative sigma‑clipping when fitting each HIL’s EW distribution with a log‑normal model. We first set a ``sigma‑limit" ($n\sigma$, where $n$ is $-2$ or $-3$; see below), perform an initial fit, and then discard all EW values below the threshold $\mu + n\sigma$. A new fit is then performed on the clipped sample, the threshold is updated, and the process repeats. Iterations will continue until the change in the fitted mean value $\mu$ falls below a pre-determined threshold of $10^{-5}$. The final converged values of $\mu$ and $\sigma$ are adopted as the best‑fit parameters. This procedure robustly excludes the low‑EW outliers and yields a stable characterization of the core EW distribution.
% We set a sigma-limit based on the EW distribution of the specific HIL. After one iteration of fitting, sources with EW values less than the sigma-limit in the distribution were discarded for the next iteration. The fitting process is considered to converge when the difference between the mean values from two consecutive fittings falls below the predetermined threshold, which is $10^{-5}$ for all HILs. 

In this analysis, we initially set the sigma‑limit for all HIL EW distributions to $–2\sigma$. This choice, while somewhat arbitrary, is proved generally effective at clipping the tail at the weak end. However, for the \Civ\ EW distribution which exhibits a substantially longer asymmetric tail, we adopted a sigma‑limit of $–3\sigma$.
% In this analysis, the sigma-limits for Si~{\sc iv}$+$O~{\sc iv}, He~{\sc ii$~\lambda$}1640, N~{\sc iii]}~$\lambda$1750 and \ciii\ were set to $-2\sigma$, and to $-3\sigma$ for C~{\sc iv$~\lambda$}1549. 
The \nv\ emission line shows a distinct EW distribution from those of other HILs. However, the EW measurement of this line may suffer from its blending with the Ly$\alpha$ emission, which is further discussed in Section~\ref{sec:measurement}. The sigma-limit for \nv\ was set to $-1.8\sigma$. 
These sigma‑limits were chosen empirically to ensure stable convergence in our iterative fitting procedure.
The final fitting results are shown in the Figure~\ref{HILs_distributions}, detailing the redshift coverage, number of objects, mean value and 3$\sigma$ range of each HIL EW distribution. The black histogram indicates the EW distribution of each HIL after applying corresponding cuts in Section~\ref{subsec:specific}; the red line represents the best-fit lognormal distribution, and the vertical dash and dash-dotted lines indicate the $2\sigma$ and $3\sigma$ ranges, respectively.

\begin{figure*}[t!]
	\includegraphics[scale=0.45]{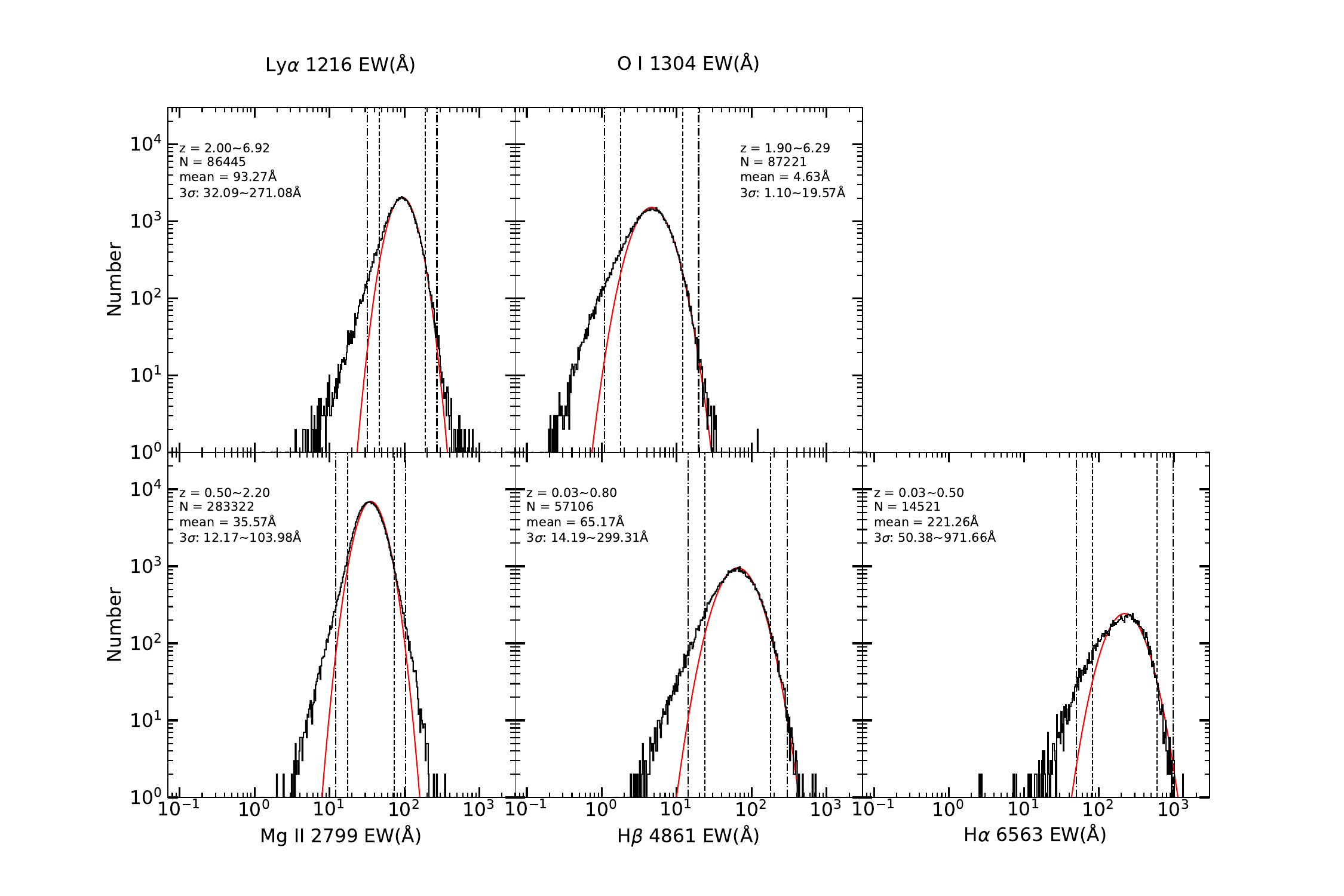}\caption{EW Distributions of LILs. The legend is the same as that in Figure~\ref{HILs_distributions}. }
	\label{LILs_distributions}
\end{figure*}

%Similarly, the redshift range of each LIL, the number of sources, $3\sigma$ range, and the mean value of fitted lognormal distribution are depicted in the upper left or right corner of each panel. Not only Ly$\alpha~\lambda$1216, all LILs show a prominent tail toward low EW values just like HILs. 

As shown in Figure~\ref{HILs_distributions}, all HILs exhibit a tail at the weak end but not at the strong end in their EW distributions, consistent with the findings of \cite{2009ApJ...699..782D} and \cite{2012ApJ...747...10W}. However, the samples in this work are significantly enlarged (by 1--2 orders of magnitude) compared to those in previous studies. Furthermore, we also fit the EW distributions of other HILs, such as \Heii\ and C~{\sc iii}], which exhibit the same behavior as \Civ. In \cite{2009ApJ...699..782D}, the EW(\Civ) distribution of high-redshift ($z>3$) quasars resulted in mean EW value of 41.9 \AA ~and the $3\sigma$ ranges of 10.0--173.7 \AA.  
The SDSS quasar sample for the EW(\Civ) distribution modeling in \cite{2012ApJ...747...10W} ($z=$1.55--4.67) has the mean EW value of 36.1 \AA\ and $3\sigma$ ranges of 10.6--123.3 \AA. The mean EW(\Civ) (41.5 \AA) of our best fitting (see Figure~\ref{HILs_distributions}) is similar to the former but slightly larger than the latter, while the $3\sigma$ range (8.2--209.5 \AA) is larger than both. The presence of a tail at the weak end of the HIL EW distribution appears to be an inherent property, and will not change even if we adopt more stringent criteria (such as SNR of the spectra).

\subsection{Low-ionization Broad Lines (LILs)}\label{LILs}
We also fit the EW distributions of LILs, as shown in Figure~\ref{LILs_distributions}, using the lognormal model and the same iterative method. The sigma-limit in the fittings is set to $-2\sigma$ for all LILs, while the predetermined threshold for converge is $10^{-5}$. The LILs studies here include Ly$\alpha$~$\lambda$1216, O~{\sc i}~$\lambda$1304, Mg~{\sc ii}~$\lambda$2799, H$\beta$~$\lambda$4861, and H$\alpha$~$\lambda$6563. Notably, in \cite{2009ApJ...699..782D} and \cite{2012ApJ...747...10W}, \Lya$+$\Nv\ was treated together as a line complex, while in WS22, these two lines were fitted separately.

As shown in Figure~\ref{LILs_distributions}, the EW histograms of the LILs exhibit a tail at the weak-line end but no significant tail at the strong-line end, except for the Mg~{\sc ii} line, which has tails at both ends. We will compare EW distributions of HILs and LILs in Section~\ref{sec:other_emission_lines}.

\subsection{Narrow Emission Lines}

\begin{figure*}[]
	\includegraphics[scale=0.50]{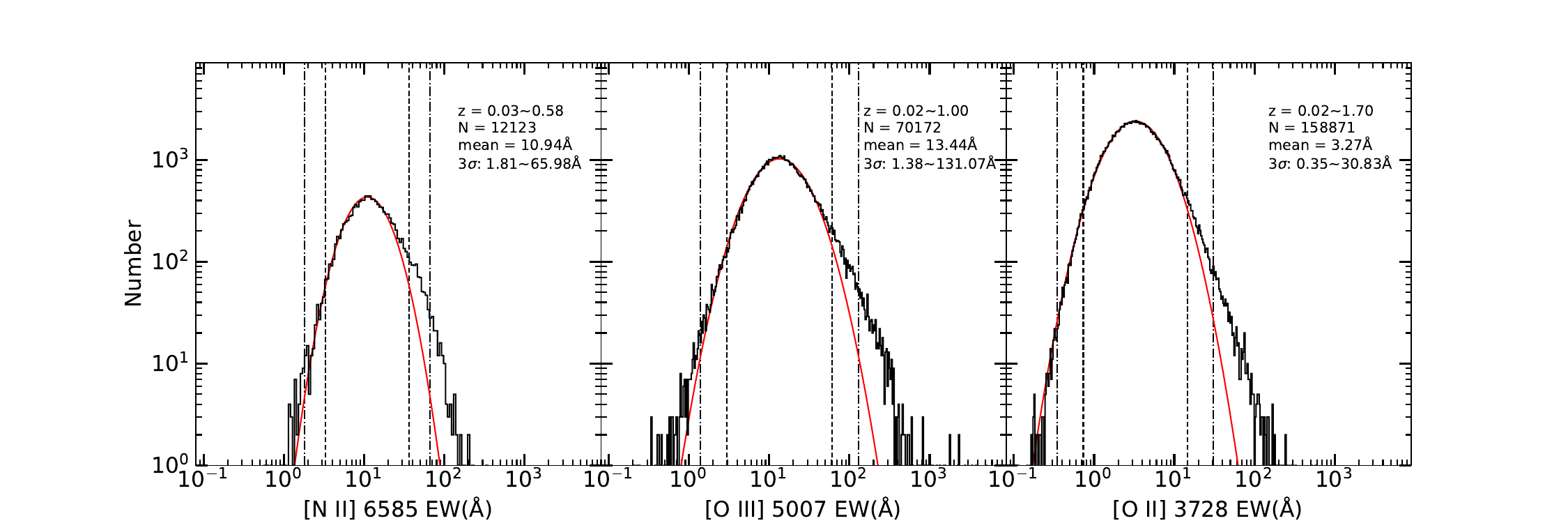}\caption{EW distributions of narrow emission lines. The legend is the same as that in Figure~\ref{HILs_distributions}. All narrow lines show a tail toward high EW values but none toward low EW values, which are exactly opposite to those of HILs.}
	\label{narrow_lines_distribution}
\end{figure*}

We continued to use the above iterative method to fit the EW distributions of the three narrow lines: [N~{\sc ii}]~$\lambda$6585, \oiii\ and [O~{\sc ii}]~$\lambda$3728. We found that the narrow lines have distinct properties from broad lines. Their EW distributions exhibit a tail (although not as substantial as those of the broad lines) at the strong-line end but not at the weak-line end, making it infeasible to discard sources with EW less than the sigma-limit. Therefore, we set positive sigma-limits to discard EW values at strong end while performing iterative fitting. These values are $2\sigma$, $2.3\sigma$, and $2.5\sigma$ for [N~{\sc ii}]~$\lambda$6585, \oiii, and [O~{\sc ii}]~$\lambda$3728, respectively. The fitting results are shown in Figure~\ref{narrow_lines_distribution}, with legends are consistent with Figure~\ref{HILs_distributions}.

Due to the limitation of SDSS spectral coverage, we cannot verify whether the narrow emission lines of \Civ-selected WLQs is generally stronger or weaker. Systematic large sample studies suggest that the narrow lines (e.g., \Oiii) of wind dominated quasar, such as WLQs, tends to be weak (e.g., \citealt{2019MNRAS.486.5335C}) and are often accompanied by strong \Oiii\ wing component (e.g., \citealt{2016ApJ...817...55S}), ruling out the possibility that WLQs contributes largely to the strong-line tail of the narrow-line EW distribution. \cite{2011MNRAS.411.2223R} and \cite{2017MNRAS.464..385B} argued that the \Oiii\ tail at higher EW values is due to projection effects. 
%where the narrow emission line is isotropic but the continuum emission is randomly oriented due to a optically thick and geometrically thin disc. Our results in Figure~\ref{narrow_lines_distribution} reconfirm this conclusion and extend it to three narrow lines.

\section{WLQs as outliers in classical relations} \label{sec:results}

\subsection{WLQs as Outliers in the \mbox{$L_{1350\rm \textup\AA}-$\rm \Civ} Blueshift Relation}\label{sec:blueshift}

\begin{figure}[t!]
	\includegraphics[scale=0.35]{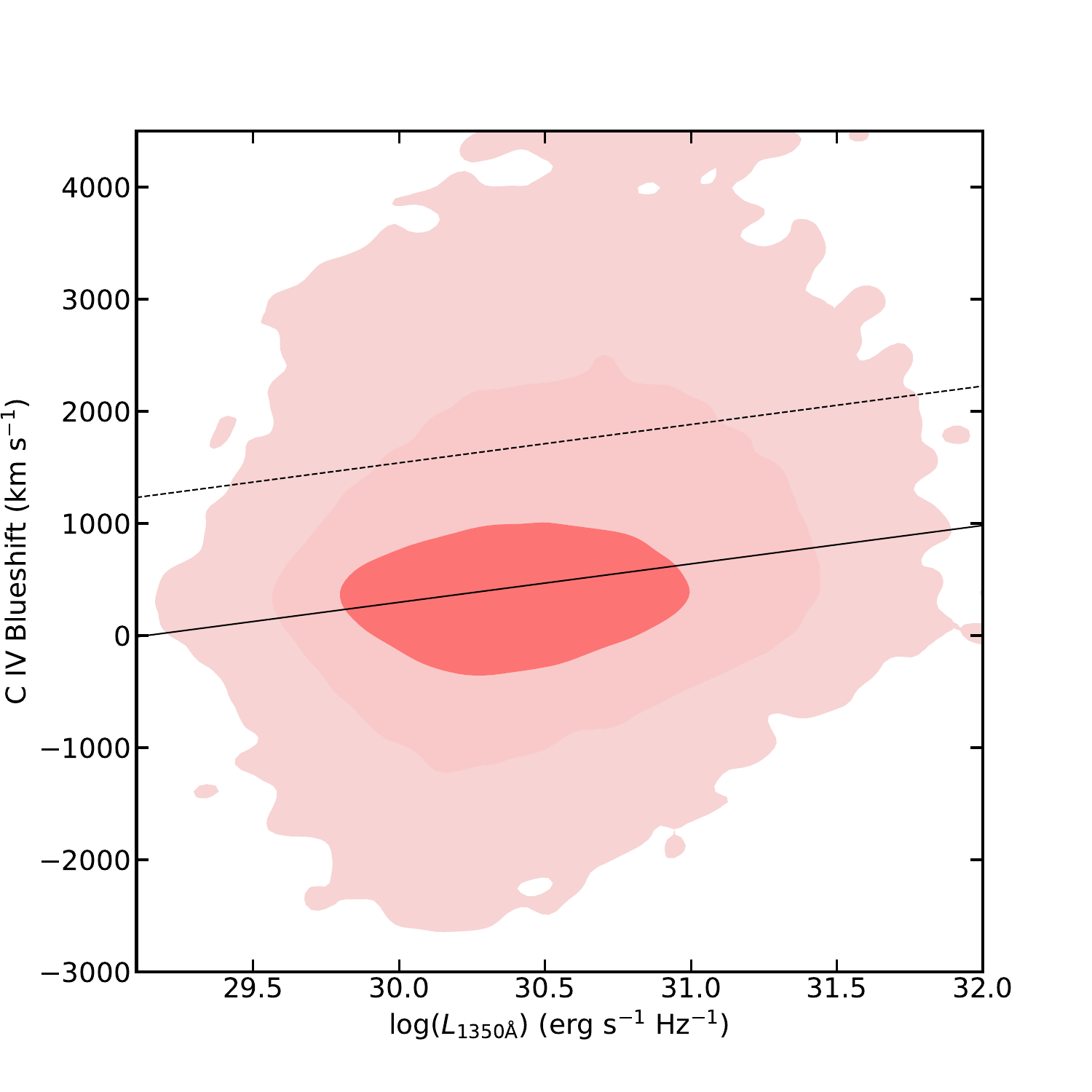}\caption{The relationship between \Civ\ blueshift and monochromatic luminosity at 1350~\AA\ in the rest frame $L_{1350\rm \textup\AA}$. The red contours from darker to lighter represent regions containing 68\%, 95\% and 99.7\% of the sources respectively. The black solid line represents the linear equation fitted by MCMC method. The black dashed line marks the boundary of outliers for the \mbox{$L_{1350\rm \textup\AA}-$\rm \Civ} blueshift relation. Region below this dashed line encompasses 95\% of the objects, while sources above the line will be considered outliers. }
	\label{L_blueshift(1350)}
\end{figure}

\begin{figure}[t]
	\includegraphics[scale=0.35]{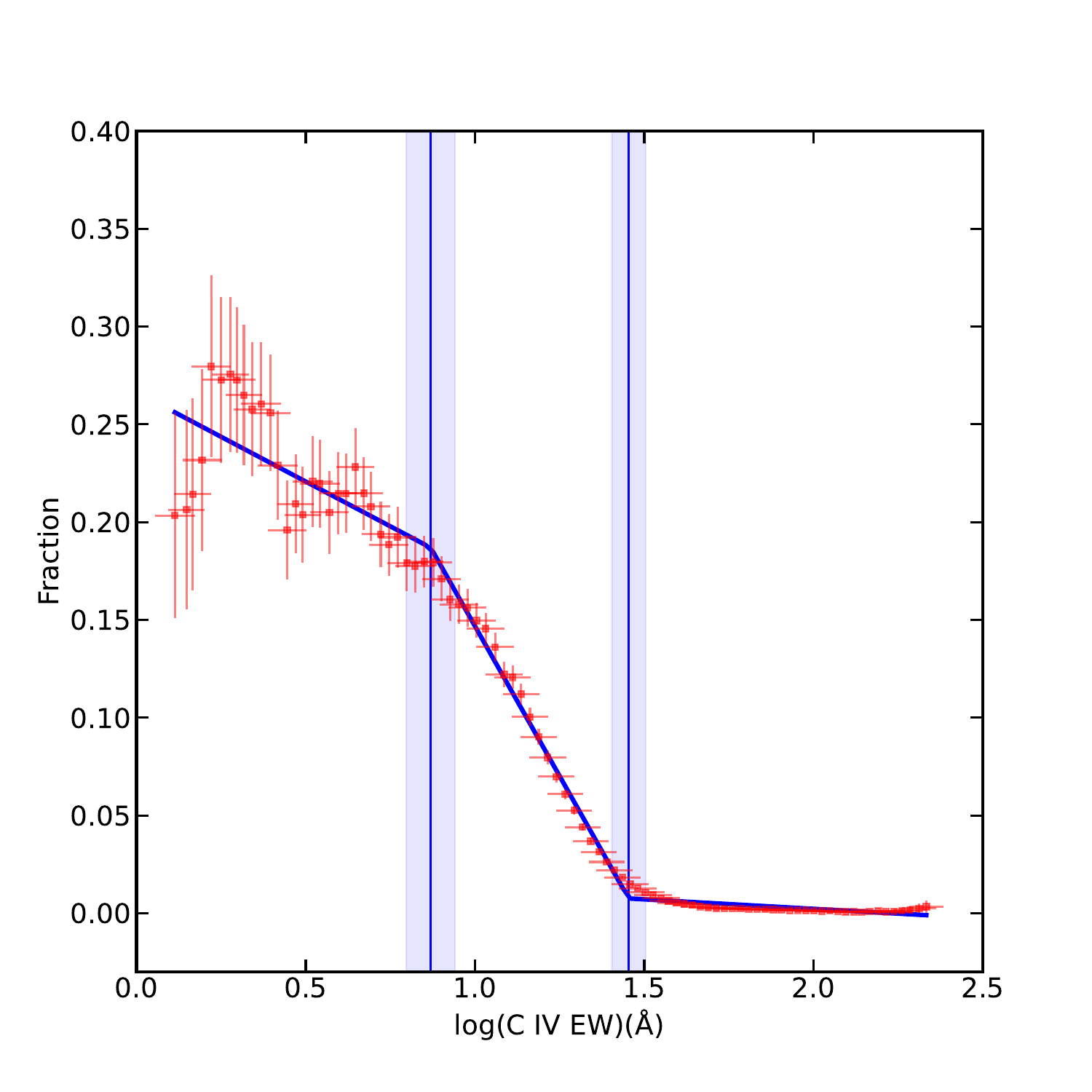}\caption{The fraction of outliers in the \mbox{$L_{1350\rm \textup\AA}-$\rm \Civ} blueshift relation as a function of EW(\Civ). We divide the sources based on their EW(\Civ) using overlapping bins to account for the limited number of sources at small EWs. The standard deviation of EW(\Civ) within each bin is shown as the x-axis error, while the y-axis error represents the binomial uncertainty. The three segments of the dark blue polyline represent the best fit linear relation in each region, while the vertical blue lines with lighter shaded areas represent location of the two breakpoints and their uncertainty ranges. }
	\label{blueshift_weakness(1350)}
\end{figure}

Some HILs such as \Heii, \Civ, and \siiv\ in the quasar spectra have strong blueshift dependence on luminosity (e.g., \citealt{1982ApJ...263...79G, 2002AJ....124....1R, 2011AJ....141..167R, 2016ApJ...831....7S}). Notably, \Civ\ shows a significant blueshift relative to its rest-frame wavelength, which becomes more prominent with increasing luminosity. We adopt the \Civ\ blueshift values from the WS22 catalog which is calculated relative to the mean systemic redshift $z_{sys}$ in the same catalog. $z_{sys}$ is derived through weighting individual redshifts $z_{line}$ by their uncertainties, with outlier values exceeding three times the median absolute deviation being rejected. 

Following \cite{2016ApJ...831....7S}, we utilize an exceptionally large sample of 132,458 sources from the WS22 catalog to investigate the \mbox{$L_{1350\rm \textup\AA}-$\rm \Civ} blueshift relationship. This sample is selected based on general criteria in Section~\ref{sec:general}, the emission-line criteria in Section~\ref{subsec:specific}, and an additional criterion requiring \mbox{$L_{1350\rm \textup\AA} > 0$}. We applied the Markov Chain Monte Carlo (MCMC) method to perform the linear regression for this relationship, which is presented in Figure~\ref{L_blueshift(1350)}. The three contours, ranging from darker to lighter colors, contain 68\%, 95\%, and 99.7\% of the sources, respectively. The solid line represents the best-fit linear relation between $\log(L_{1350\rm \textup\AA}/{\rm erg\ s^{-1}\ Hz^{-1}})$ and \Civ\ blueshift (km~s$^{-1}$), with the slope and intercept as $342.6\pm5.8$ and $-9963.4\pm146.3$, respectively. We calculated the Pearson correlation coefficient for this relationship $r=0.192$. There exists a significant scatter between luminosity and \Civ\ blueshift, and the distribution of \Civ\ blueshift residuals as a function of luminosity appears asymmetric. A small subset of sources exhibits notably strong \Civ\ blueshifts, with values exceeding approximately 2000 km~s$^{-1}$, as shown in Figure~\ref{L_blueshift(1350)}. To better identify outliers with large \Civ\ blueshift values, we define a range around the \Civ\ blueshift residuals, centered on zero, that encompasses approximately 95\% of the sources. Objects outside this range are classified as outliers. In Figure~\ref{L_blueshift(1350)}, we plot the boundary of this range for the positive residuals (black dashed line). 

The correlation, with the highest velocity associated with the most luminous quasars, is consistent with the expectations for the radiatively-driven outflow scenario. Noting that WLQs generally have stronger \Civ\ blueshift \citep{2015ApJ...805..122L} while being located at the weak-line end of the EW(\Civ) distribution, we created Figure~\ref{blueshift_weakness(1350)} to check whether these outliers in the \mbox{$L_{1350\rm \textup\AA}-$\rm \Civ} blueshift relationship are mainly WLQs. 
% As described above, sources above the dashed line in Figure~\ref{L_blueshift(1350)} are considered outliers in the \mbox{$L_{1350\rm \textup\AA}-$\rm \Civ} blueshift relationship, exhibiting larger \Civ\ blueshift values than expected for their luminosity. 
Examining these outliers,  we find that the their EW(\Civ) values are more likely to be small, positioning them in the lower right corner in the $\log$(EW(\Civ))-\Civ\ blueshift parameter space (e.g., \citealt{2011AJ....141..167R, 2020ApJ...899...96R, 2022ApJ...931..154R}).

To systematically track the dependence of outlier prevalence on \Civ\ strength, we made Figure~\ref{blueshift_weakness(1350)} to observe how the fraction of outliers varies with EW(\Civ). We divided the typical range of EW(\Civ) into 100 logarithmically spaced bins with partial overlap and then calculated the fraction of outliers in each bin. The red points with error bars represent the fraction within each bin, showing a clear trend where these fractions decline as a function of EW(\Civ). To quantify this trend, we applied a piecewise regression analysis using the \mytt{piecewise-regression} package.\footnote{https://github.com/chasmani/piecewise-regression} The results show that for the \mbox{$L_{1350\rm \textup\AA}-$\rm \Civ} blueshift relationship, the dependence of the outlier fraction upon the EW(\Civ)  can be segmented into three distinct regions, reflecting systematic variations of emission-line kinematics within the quasar sample. When EW(\Civ) is greater than $28.4_{-3.0}^{+3.4}~\rm \AA$, the fraction of outliers remains flat around 0, while for the EW(\Civ) range of $7.4_{-1.1}^{+1.3}$ to $28.4_{-3.0}^{+3.4}~\rm \AA$, the fraction of outliers increases rapidly as EW(\Civ) decreasing. However, when the EW(\Civ) is less than $7.4_{-1.1}^{+1.3}~\rm\AA$, the increasing trend is interrupted, and the outlier fraction fluctuates between $\sim20\%$ and $30\%$. %rate of increase is significantly reduced or the outlier fraction even drops. , so that it appears as a bump at $7.4_{-1.1}^{+1.3}~\rm\AA$. 

Figure~\ref{blueshift_weakness(1350)} is another illustration of the inner connections between the larger blueshift and smaller EW values of the \Civ\ emission line, which indicates the same physical mechanisms. This has been characterized by the so-called ``\Civ\ Distance" parameters \citep{2020ApJ...899...96R,2022ApJ...931..154R}, which is connected to the Eddington ratio of quasar accretion. The strong \Civ\ blueshift and weak \Civ\ line strength are often manifested simultaneously in WLQs (e.g. \citealt{2011ApJ...736...28W, 2012ApJ...747...10W, 2015ApJ...805..122L}). Understanding the driving mechanisms behind these characteristics is crucial for comprehending the formation and structure of WLQs. According to the accretion disk-wind model, higher accretion rate will generally lead to strong outflows, resulting in strong blueshift of the wind-dominated \Civ\ emission line. This can also explain the different behaviors between the \Civ\ line and the disk-dominated lines such as \Hb\ and \Mgii\  (e.g., \citealt{2011AJ....141..167R,2015ApJ...805..123P}). 
%WLQs are a type of quasar whose \civ\ line emission is dominated by disk wind,  with their \civ\ emission in the disk wind and disk components is greatly suppressed \citep{2015ApJ...805..123P}. 

\subsection{WLQs as Outliers in the Baldwin Effect}\label{sec:BE}

The Baldwin effect, which denotes the inverse correlation between the emission line strength (specifically the broad \civ\ emission line) and the continuum luminosity, has been discovered since 1977 \citep{1977ApJ...214..679B}. Various explanations have been proposed, including the Eddington ratio \citep{2004MNRAS.350L..31B,2015ApJ...805..124S}, SMBH mass \citep{2008MNRAS.389.1703X}, and metallicity \citep{2004ApJ...608..136W}. Despite these efforts, no definitive theories have gained consensus. 

In this study, we continue to use the subsample described in Section~\ref{sec:blueshift} to examine the Baldwin effect. This subsample exceeds previous studies in terms of sample size (e.g., \citealt{2001ApJ...556..727G, 2002ApJ...581..912D, 2004ApJ...608..136W, 2008MNRAS.389.1703X, 2011AJ....141..167R, 2012MNRAS.427.2881B}). Our focus in this section is not to pinpoint the physical driver of the Baldwin effect, but to highlight the anomalous behaviors of WLQs regarding this effect.

\begin{figure}[t]
	\includegraphics[scale=0.35]{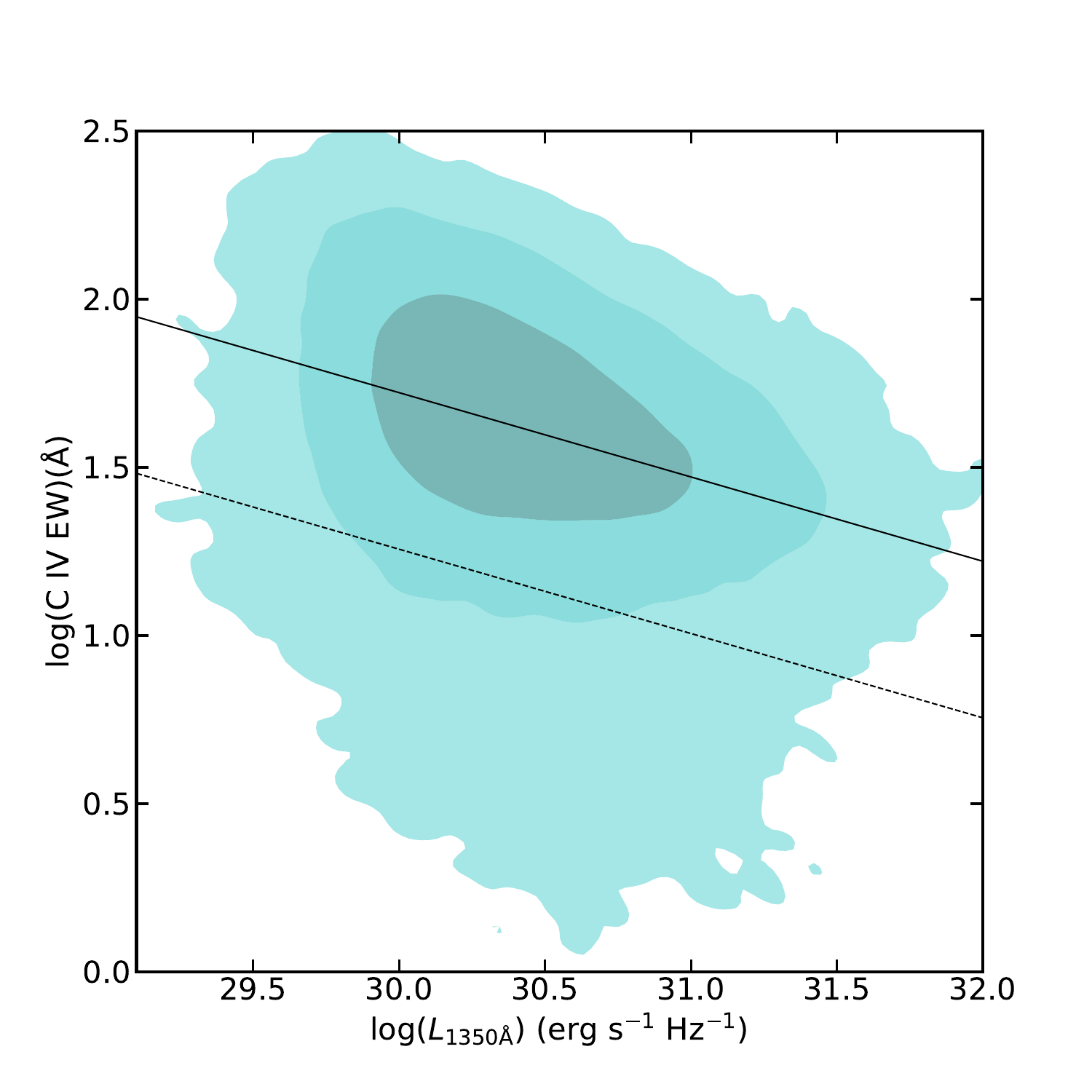}\caption{Baldwin effect for quasars from the DR16 subsample. The cyan contours, from darker to lighter, represent regions containing 68\%, 95\%, and 99.7\% of the sources, respectively. The solid line indicates the MCMC-fitted linear equation, while the dashed line marks the boundary encompassing approximately 95\% of the sources. Objects below this dashed line are considered as outliers in the Baldwin effect. The scatter in the Baldwin effect is smaller than that in the \mbox{$L_{1350\rm \textup\AA}$}-\Civ\ blueshift relation.
	}
	\label{BE_L1350}
\end{figure}

\begin{figure}[t]
	\includegraphics[scale=0.35]{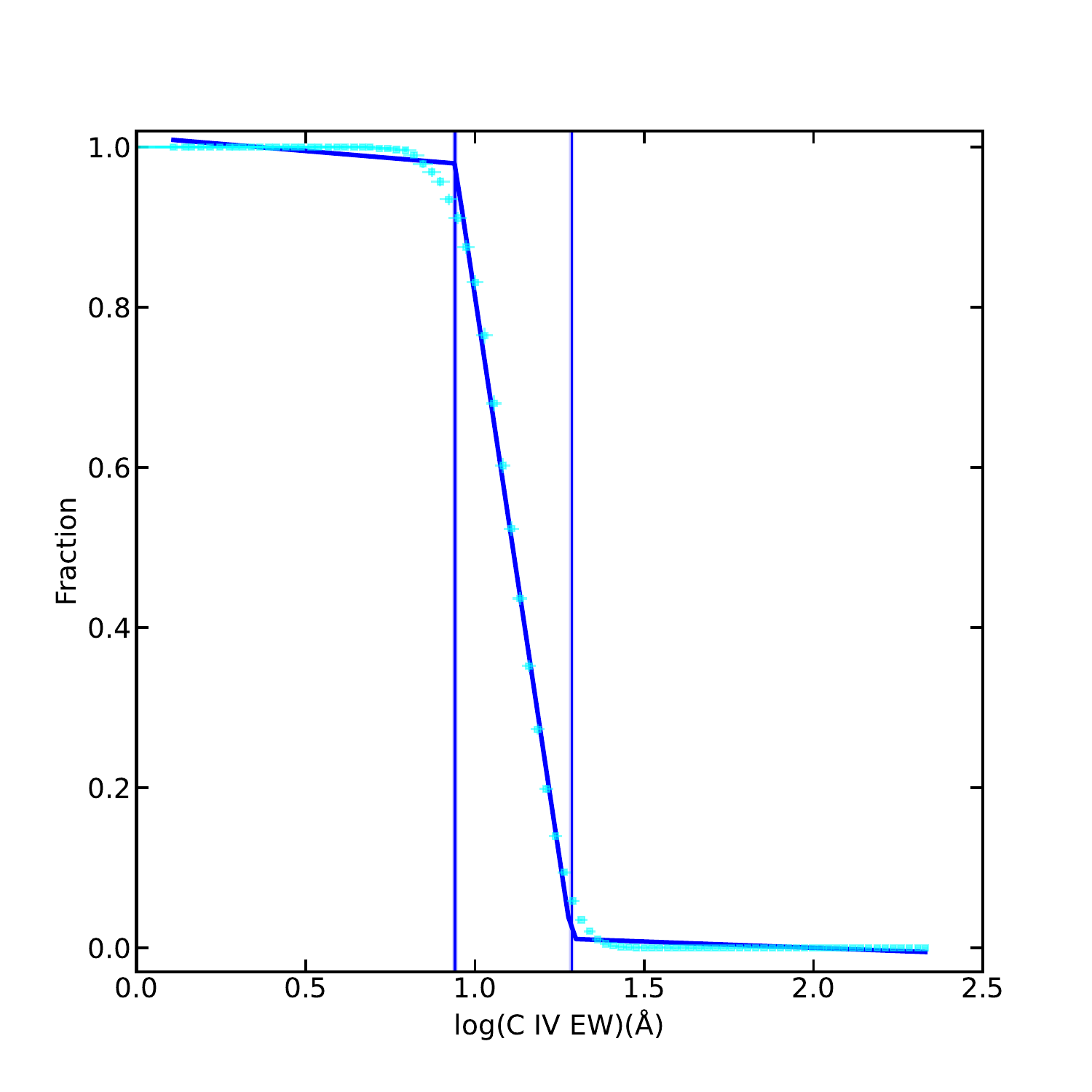}\caption{The fraction of outliers in the Baldwin effect as a function of EW(\Civ). The method adopted here is consistent with that in Figure \ref{blueshift_weakness(1350)}. The cyan points with error bars represent these fractions. The dark blue polyline segments indicate the best fit for each region, with the vertical blue lines marking the locations of the breakpoints. The light blue shaded areas represent the uncertainties of the two breakpoints, located at $8.7^{+0.1}_{-0.1}$~\AA\ and $19.3^{+0.3}_{-0.3}$~\AA, respectively. }
	\label{EW_weakness}
\end{figure}

We observe that the Baldwin effect itself exhibits considerable scatters, which are particularly pronounced in quasars with extreme properties, such as WLQs. For instance, \cite{2004ApJ...611..107L} analyzed narrow line Seyfert 1 galaxies (NLS1s, viewed as low redshift and low SMBH mass counterparts of WLQs, see more discussion in Section~\ref{subsec:ledd}) in the Baldwin diagrams for UV emission lines and found that NLS1s are systematically offset to lower continuum luminosity at a given EW(\Civ), relative to typical broad-line AGNs. NLS1s' characteristics overall place them at one extreme of the eigenvector 1 sequence \citep{1992ApJS...80..109B} and there is considerable belief that they are in a high accretion state with large Eddington ratio $\lambda_{\rm Edd}$. Similarly, \cite{2015ApJ...805..124S} reported substantial scatters in the $\lambda_{\rm Edd}$-modified Baldwin effect among WLQs, typically locating them in the lower right of the log$\lambda_{\rm Edd}$-log EW(\Civ) parameter space, indicating lower EW(\Civ) than expected from the modified Baldwin effect given their Eddington ratio. After excluding 12 radio loud quasars, 7 BAL quasars and 11 WLQs in high-z subsample, the Baldwin effect becomes stronger in \cite{2016MNRAS.462..966G}, which shows that WLQs often exist as outliers in the Baldwin effect, although their absolute number is small. In \cite{2023ApJ...950...97H}, their WLQs deviate from the best-fit model by a mean of 3.4$\sigma$, with a range in deviation from 1.08$\sigma$ to 8.02$\sigma$.

In order to quantify the deviation of WLQs from the Baldwin effect, we define outliers using the same method outlined in Section~\ref{sec:blueshift}. Figure~\ref{BE_L1350} illustrates the Baldwin effect between $L_{1350\rm \textup\AA}$ and EW(\Civ) for the same subsample. The solid line represents the MCMC-fitted linear correlation. The slope and intercept of the fitted equation are $-28.37\pm0.19$ and $911.38\pm6.03$, respectively. The Pearson correlation coefficient between \mbox{$L_{1350\rm \textup\AA}$} and log EW(\Civ) is $-0.357$. The dashed line marks the boundary encompassing approximately 95\% of the sources with negative residuals. Objects below this line are outliers for which the expected EW(\Civ) from the Baldwin effect is not achieved at their respective luminosities. In Figure~\ref{EW_weakness}, we divided the EW(\Civ) into 100 logarithmically spaced bins with partial overlap and calculated the fraction of outliers in each bin. We selected  \mbox{$L_{1350\rm \textup\AA}$-\rm EW(\Civ)} relation for determining outliers due to its strong correlation and relatively less scatter compared to the monochromatic luminosity at other wavelengths. The fraction of outliers, as illustrated in Figure~\ref{EW_weakness}, is distinctly divided into three segments: for EW(\Civ) $\textless$ $8.7^{+0.1}_{-0.1}$~\AA, the fraction remains at around 100\%, indicating that sources with EW(\Civ) below  this threshold are almost entirely outliers in the Baldwin effect; for the part where $8.7^{+0.1}_{-0.1}~$\AA$~ \textless$ EW(\Civ) $\textless$ $19.3^{+0.3}_{-0.3}$~\AA, the fraction of outliers drops steadily from about 100\% to 0\%, representing a transition zone between normal quasars and the outliers; for the objects with EW(\Civ) ~$\textgreater$ $19.3^{+0.3}_{-0.3}~$\AA , the fraction of outliers stabilizes at around 0\%. Figure~\ref{EW_weakness} demonstrates that the Baldwin effect only holds over an intermediate EW range. For for quasars with extreme properties, such as WLQs, the Baldwin effect alone may not be able to explain the weakness of the \Civ\ emission line. Instead, additional mechanisms, such as the blocking of ionizing photons by the shielding gas, could also be contributing. 
%when EW(\Civ) is very small, the deviation cannot be attributed solely to a softer ionizing continuum (e.g., \citealt{1992MNRAS.254...15N, 1993ApJ...415..517Z}) but instead suggests additional mechanisms such as absorption of high energy photons by shielding gas. }
%Overall, this shows that for quasars with extreme properties, such as WLQs, the Baldwin effect alone may not be able to explain the weakness of the emission line. 

\subsection{WLQs as Outliers in the \mbox{$\log L_{2500\textup\AA}-\alpha_{\rm ox}$} Relation}\label{sec:ox}

\begin{figure}[tp]
	\includegraphics[scale=0.35]{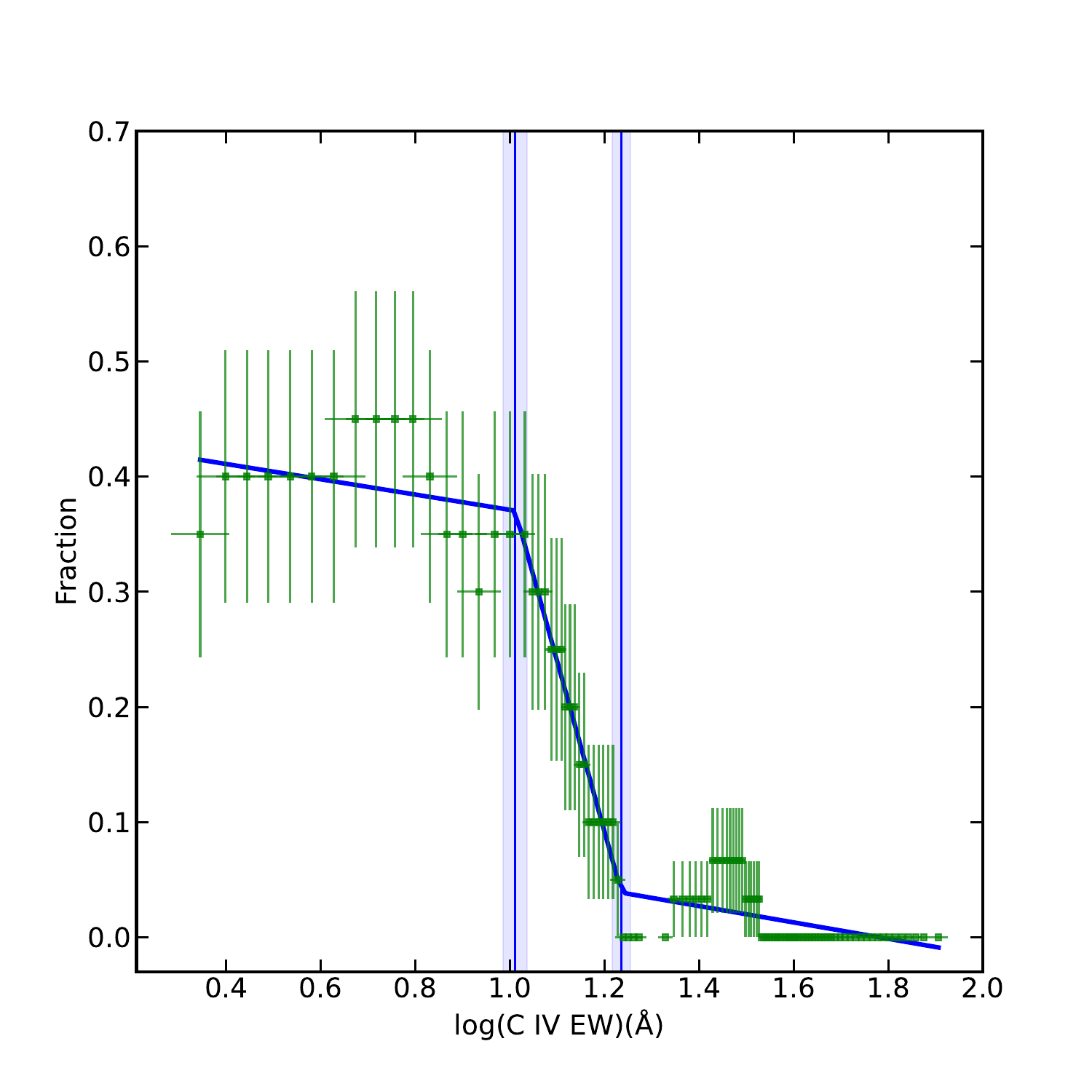}\caption{The fraction of X-ray weak sources (by a factor of $\sim3.3$ compared to typical quasars with similar UV luminosity) as a function of EW(\Civ), represented by the green points with error bars. The three segments of dark blue polyline represent the best fit linear relation in each region. The vertical blue lines and lighter shaded areas represent location of the breakpoints between the three segments and their error bars. }
	\label{x_weakness}
\end{figure}

WLQs have garnered significant attention as outliers in the  \mbox{\rm $\log L_{2500\textup\AA}-\alpha_{\rm ox}$} relationship, primarily because nearly half of this population exhibits exceptionally weak X-ray emission \citep{2011ApJ...736...28W, 2012ApJ...747...10W, 2015ApJ...805..122L, 2018MNRAS.480.5184N, 2022MNRAS.511.5251N}. \cite{2018MNRAS.480.5184N} systematically studied the relationship between the fraction of X-ray weak quasars and EW(\Civ), finding a strong dependence: the fraction of X-ray weak quasars decreases by $\sim$13 times as the increase of EW(\Civ). In their study, the fraction of X-ray weak quasars is almost stable below 10\% when EW(\Civ) \textgreater~20~\AA. However, when EW(\Civ) is reduced to $\sim$10--20~\AA, the fraction increases sharply from less than 10\% to nearly half, then stabilizes again when EW(\Civ) \textless~10~\AA. Our work in this section aims to use the \mytt{piecewise-regression} package to precisely determine the EW(\Civ) values at the two breakpoints and provide the associated uncertainties.

The ``representative sample" in \cite{2018MNRAS.480.5184N, 2022MNRAS.511.5251N} is the largest and least biased WLQ sample with X-ray observations to date. The WLQ sample in Figure~\ref{x_weakness} closely matches the representative sample in \cite{2018MNRAS.480.5184N}, thus we also use the ``sample B" in \cite{2008ApJ...685..773G} as typical quasars. 
% For the first 40 sources, we set the run step and run window to 1 and 20, respectively. After the 40th source, we adjust the run step and run window to 2 and 30, respectively, with error bars representing 1$\sigma$ confidence intervals obtained through bootstrapping. 
%Given the description of the fraction of X-ray weak quasars and the dependence trend of EW(\Civ) above, we will not elaborate further here. 
Following method in \cite{2018MNRAS.480.5184N}, we analyzed the X-ray weakness ($\Delta\rm\alpha_{\rm ox}\textless -0.2$ relative to the expectations based on the \mbox{\rm $\log L_{2500\textup\AA}-\alpha_{\rm ox}$} relation in \citealt{2007ApJ...665.1004J}) dependence of EW(\Civ) using an adaptive sliding window approach. Sources were sorted by ascending EW(\Civ), with initial calculations using a 20-source window moving in 1-source steps through the first 40 sources. When the window's right edge reached the 40th source, we expanded the window to 30 sources with 2-source stepping to improve statistics. At each position, we computed both the mean EW(\Civ) and the fraction of X-ray weak quasars with error bars representing 1$\sigma$ confidence intervals.
Using the \mytt{piecewise-regression} package, we accurately measure the first and second breakpoints at EW(\Civ) $=10.3^{+0.6}_{-0.6}$~\AA\ and EW(\Civ) $=17.2^{+0.8}_{-0.7}$~\AA, respectively. The first breakpoint is basically consistent with the estimation in \cite{2018MNRAS.480.5184N}, while the second breakpoint is slightly smaller.

From the perspective of the shielding gas model, the dependence between the fraction of X-ray weak quasars and the EW values of HILs suggests that the more severely shielded the quasar is (indicated by higher relative height and column density of the shielding gas), the fewer high-energy photons from inner disk and corona reach the BLR. This results in weaker HILs (most notably  C~{\sc iv}) in the optical/UV spectra of WLQs. For X-ray emission in WLQs, if our line of sight intersects the X-ray absorbing shielding gas, X-ray weak WLQs are observed, characterized by $\Delta\rm\alpha_{\rm ox}\textless-0.2$; Conversely, if the line of sight does not go through the shielding gas, the WLQs are observed as X-ray normal. Consequently, the increasing degree of obscuration leads to a correlation between the fraction of X-ray weak quasars and EW(\Civ).
%In contrast, X-ray emission in normal quasars remains unaffected by viewing angle, as no such obscuration is present. 

% Given that \Heii\ and \Civ\ have similar ionization energies, their emission line properties are expected to be similar. Indeed, the \heii\ and \civ\ emission lines often show high blueshifts in the same quasar, and their blueshift trends with luminosity are synchronized \citep{2016ApJ...831....7S}. This indicates their close ionization locations in the BLR. Thus, the shielding gas model naturally explains the similar weak strength of HILs due to the blocking of high-energy photons. Unfortunately, we are currently unable to analyze the dependence between the fraction of X-ray weak quasars and \Heii\ EW as we have done for EW(\Civ), mainly because \heii\ is not as prominent in the quasar UV spectrum and nearly invisible in WLQ spectra. Future studies with higher-quality spectra and more advanced measurement techniques may help address these challenges.

\subsection{Definition of WLQs}\label{sec:definition}

As discussed in Sections~\ref{sec:blueshift}--\ref{sec:ox}, WLQs consistently appear as outliers in the quasar UV luminosity$-$\Civ\ blueshift relation, the Baldwin effect, and the UV luminosity$-\alpha_{\rm ox}$ relationship. This consistent behavior suggests that the accretion state of WLQs may differ significantly from that of normal quasars. It is generally believed that WLQs have relatively high Eddington ratios (close to or greater than 1; e.g., \citealt{2015ApJ...805..122L,2018ApJ...865...92M}), leading to a transition in their (inner) accretion disk from a standard thin disk to a slim disk. In this sense, we assert that WLQs represent a distinct population at unique accretion states, with unusual X-ray and optical/UV spectroscopic properties. 
% rather than the extreme end of normal quasars.

As stated in the introduction, existing definitions of WLQs with thresholds like EW(Ly$\alpha$+N~{\sc v}) \textless~15.4~\AA\ and/or EW(\Civ) \textless~10~\AA\ are statistically driven, which do not fully capture their underlying physical nature. Figures~\ref{blueshift_weakness(1350)}, \ref{EW_weakness}, and \ref{x_weakness} illustrate that the fraction of outliers in these three relationships is highly dependent on EW(\Civ). They all follow the same trend that the outlier fraction increases sharply when EW(\Civ) is in the $\sim$10--20~\AA\ range, while remains constant or increases slowly when EW(\Civ) $\lesssim$~10~\AA. Based on this, we propose to categorize quasars into three populations using the two EW(\Civ) breakpoints. We calculate the value of each breakpoint as the average of the regression results in Sections~\ref{sec:blueshift}--\ref{sec:ox} weighted by their individual uncertainties, as illustrated in Figures~\ref{blueshift_weakness(1350)}, \ref{EW_weakness}, and \ref{x_weakness}. The first and second breakpoints are averaged at EW(\Civ) $=8.9\pm0.2$~\AA\ and EW(\Civ) $=19.3\pm0.3$~\AA, respectively. In conclusion, we define quasars with EW(\Civ)~\textless~$8.9\pm0.2$~\AA\ as WLQs, representing a population generally exhibits enhanced \Civ\ blueshift, deviation from the canonical Baldwin effect, and exceptionally high fraction ($\sim50\%$) of X-ray weak objects. Quasars with EW(\Civ)~\textgreater~$19.3\pm0.3$~\AA\ are considered normal quasars with typical quasar characteristics. Quasars with EW(\Civ) between $8.9\pm0.2$~\AA\ and $19.3\pm0.3$~\AA\ are classified as bridge quasars, representing a transition state between normal quasars and WLQs. 

Under the assumption that the accretion disk of a WLQ with high accretion state has its inner part puffed up (see more discussion in Section \ref{subsec:ledd}), we thus speculate that the two EW(\Civ) thresholds adopted in our WLQ definition, namely 19.3~\AA\ and 8.9~\AA, may reflect the changing ionization state of the BLR as the accretion disk transitions from a standard thin disk (normal quasars) to a slim disk (WLQs). The bridge quasars having EW(\Civ) values between these two thresholds, represent a transition phase between the two accretion disk states. 
% From a quasar population perspective, sources with EW(\Civ) \textgreater~19.3~\AA\ likely correspond to quasars dominated by a standard thin accretion disk. As the accretion rate increases toward $\sim0.2 \pm 0.1$ (e.g., \citealt{2017A&A...608A.122S}), the system may enter a transitional state between standard and slim disks, during which the inner disk becomes increasingly geometrically thick, progressively shielding the BLR from ionizing radiation as the shielding gas model described. Once the accretion rate becomes sufficiently high, the suppression of high-energy photons and the BLR ionization state may reach a saturation point, resulting in extremely weak \Civ emission (EW \textless~ 8.9 ~\AA\ ). In this regime, WLQs significantly deviate from classical quasar relations and their accretion disk is likely dominated by powerful winds. 
This smooth transition, rather than a sudden phase change, appears more consistent with the underlying physical processes. Furthermore, because of the existence of this transition phase, WLQs would not appear as a disjoint population to normal quasars in the line parameter space, as stated in \citet{2023ApJ...950...97H}.

We have tested that even if our definition of outliers shown in Figures~\ref{L_blueshift(1350)} and \ref{BE_L1350} varies, the relationship between the outlier fraction and EW(\Civ) persists. Therefore, our conclusions that WLQs and bridge quasars represent distinct populations from normal quasars remain robust.  
%and are not influenced by how the outliers are classified.

This new definition resolves a longstanding issue by providing a classification that reflects the underlying physical nature of WLQs. With the precise criteria, we can more effectively study the continuum spectrum and emission line properties of WLQs and normal quasars over a wide range of wavelength coverage. Furthermore, applying this definition to the DR16Q catalog significantly expands the number of WLQs. For the redshift range of $z=1.45$--5.41 where the \Civ\ emission is covered by the SDSS spectroscopy, the number of WLQs and bridge quasars in DR16Q catalog after applying selection criteria are 3236 ($1.8\pm0.3\%$) and 12562 ($7.0\pm0.6\%$), respectively. Although these fractions are small, they provide a substantial sample size for detailed study.

\section{The Physical Mechanism of Emission Line Weakness}\label{sec:other_emission_lines}

% \begin{figure}[tp]
	%     \includegraphics[scale=0.35]{luminosity.pdf}\caption{The \mbox{$L_{1350\rm \textup\AA}$} distribution of WLQs and  and normal quasars. Blue and red histogram represent the \mbox{$L_{1350\rm \textup\AA}$} distribution of WLQs and normal quasars, blue and red dash line represent their median \mbox{$L_{1350\rm \textup\AA}$}, which is $30.6^{+0.1}_{-0.1}$ and $30.4^{+0.1}_{-0.1}$(68\% confidence interval), corresponding EW(\Civ) is $37.6^{+0.2}_{-0.3}$ and $42.3^{+0.1}_{-0.1}$, respectively. The \mbox{$L_{1350\rm \textup\AA}$} distribution of normal quasars can not be fitted well cause luminous quasars are easier to be observed which lead that it is not a well Gaussian distribution. This figure shows that Baldwin effect in WLQs is negligible for the weakness of emission lines.}
	%     \label{luminosity}
	% \end{figure}

\begin{figure*}[]
	\includegraphics[scale=0.50]{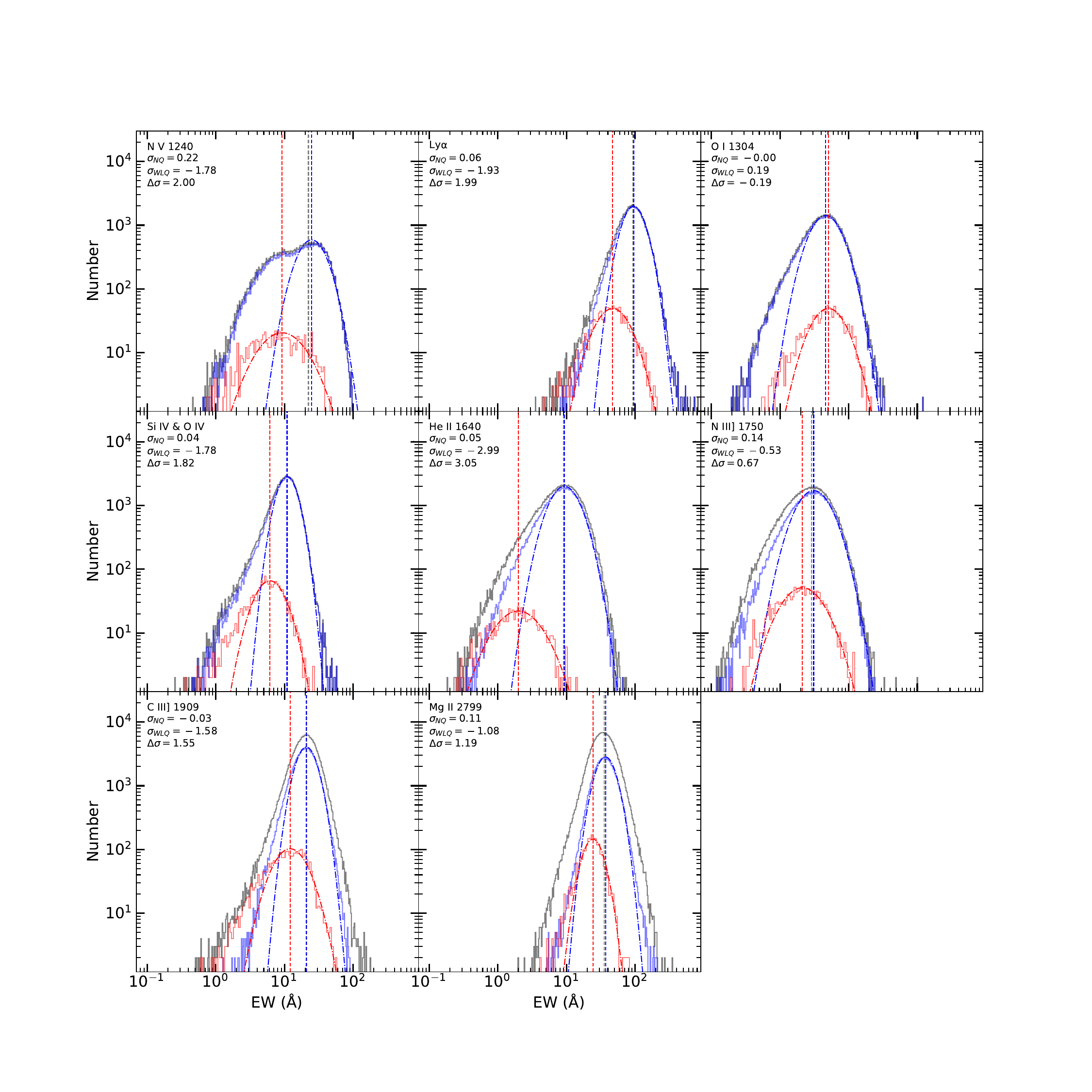}
    \caption{Comparing the EW distribution of a variety of  emission lines for WLQs and normal quasars. The red, blue and gray histograms in each panel represent the EW distributions of WLQs, normal quasars, and the full quasar population, respectively. The red, blue, and gray vertical lines correspond to the median EW values of WLQs, normal quasars, and the full quasar sample, obtained by fitting their respective distributions using lognormal models. In the upper left corner of each panel, we report $\sigma_{\rm NQ}$, $\rm\sigma_{WLQ}$, and $\Delta\sigma$ for each emission line, with $\Delta\sigma$ representing the difference between the $\sigma$ values of normal quasars and WLQs. The attenuation factor ($\Delta\sigma$) serves as a measure of how much weaker the emission lines of WLQs compared to those of normal quasars.}
	\label{WLQs_oth_lin}
\end{figure*}

In Section~\ref{sec:definition}, we established definitions for WLQs and bridge quasars based on the relationship between EW(\Civ) and the fraction of outliers, providing clear classification criteria. Applying the WLQ definition to the DR16Q catalog allows for a quantitative study of how various optical/UV emission lines of WLQs differ from those of normal quasars. Our analyses are restricted to redshifts greater than 1.45 to ensure that the \Civ\ emission line is presented in the SDSS spectra.

The aim of this section is to assess the attenuation factor for a variety of optical/UV emission lines in the WLQ spectra compared to those of normal quasars. We select WLQ and normal quasar samples from the DR16Q catalog that meet the criteria outlined in Section~\ref{sec:sample} and conform to the definitions of WLQs and normal quasars in Section~\ref{sec:definition}. The median \mbox{$L_{1350\rm \textup\AA}$} values for WLQ and normal quasar samples are $30.6^{+0.1}_{-0.1}$ and $30.4^{+0.1}_{-0.1}$, respectively. Since the luminosity of WLQs is consistent within the errors with that of normal quasars, the Baldwin effect on other emission lines will be neglected in the following discussions. The EW distributions of various emission lines for WLQs and normal quasars are compared in Figure~\ref{WLQs_oth_lin}. In each panel, the red and blue histograms represent the EW distributions of the WLQ and normal quasar samples, respectively, while the gray histograms reflect the EW distributions for the full quasar population (same as the black lines in Figures~\ref{HILs_distributions} and \ref{LILs_distributions}). Using the same methodology described in Section~\ref{HILs}, we fit the EW distributions for WLQs and normal quasars using lognormal models for each emission line, marked by red and blue dash-dotted lines in Figure~\ref{WLQs_oth_lin}. The red, blue, and gray vertical lines indicate the mean EW values  from the best-fit models for the WLQ, normal quasar, and full quasar EW distributions, respectively. We can see that the mean EW values of the normal quasar sample nearly align with those of the full quasar population, whereas the mean EW values for WLQs deviate in varying degrees. 

For each emission line, we quantify the deviation with the parameter $\sigma_{\rm WLQ}$, defined as the difference between the mean EW values for WLQs and for the full quasar population, normalized by the standard deviation of the best-fit lognormal model for the full quasar population. The parameter for normal quasars $\sigma_{\rm NQ}$ is similarly defined. 
Then, we use the difference between $\sigma_{\rm NQ}$ and $\sigma_{\rm WLQ}$ to quantify the attenuation factor for each specific emission line, represented by $\Delta\sigma$. These values are displayed in the upper left corner of each panel in Figure~\ref{WLQs_oth_lin}. 
% We have also tried to directly use the mean EW value of fitted distribution to represent the overall behavior of WLQs and normal quasars, and the results are basically consistent with the above results, this is because emission line EW of WLQs and normal quasars basically conform to the lognormal distribution plus a tail at weak end (see Figure~\ref{HILs_distributions} and Figure~\ref{LILs_distributions}), and sources near the median or mean account for the majority.

\begin{figure}[tp]
	\includegraphics[scale=0.35]{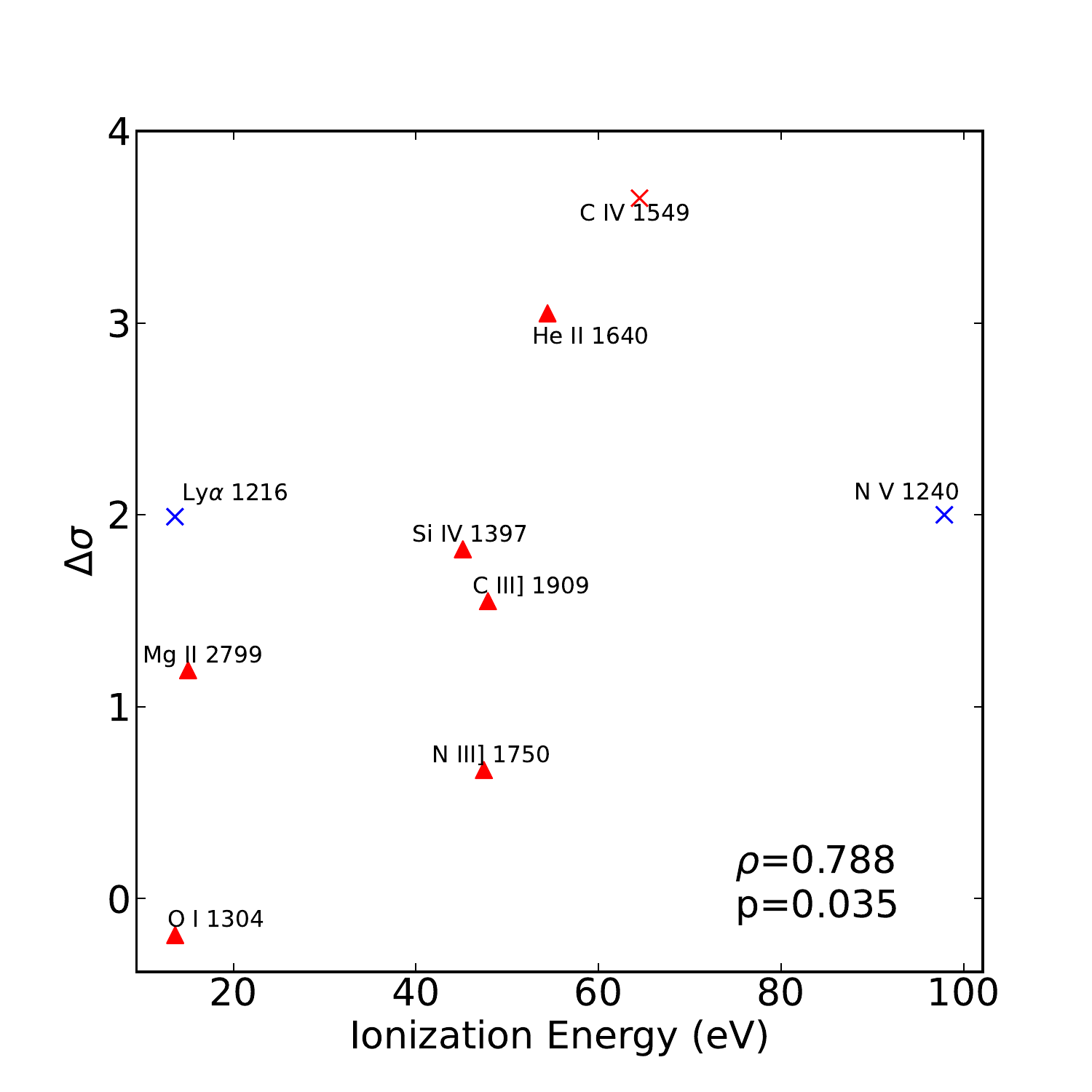}\caption{The relationship between emission line attenuation factor (depicted by $\Delta\sigma$) and the ionization energy of the corresponding ion (retrieved from \citealt{2001AJ....122..549V}). \lya\ and \nv\ are represented by blue crosses, \civ\ is represented by a red cross, while other emission lines are represented by red triangles. The correlation coefficient $\rho$ and $p$-value are given in the lower right. Note that \lya\ and \nv\ are excluded when calculating correlation coefficient.}
	\label{ionization_energy}
\end{figure}

Comparing the $\Delta\sigma$ values of different emission lines shows that the attenuation factor of a specific emission line appears related to whether it is a high or low ionization line. The most attenuated line is \heii, with $\Delta\sigma=3.05$, while the least attenuated is \oi, with $\Delta\sigma=-0.19$. The negative $\Delta\sigma$ value for \oi\ may result from intrinsic scatter of the statistical distribution, indicating little or no attenuation for this line in WLQs. Figure~\ref{ionization_energy} plots the attenuation factor (depicted by $\Delta\sigma$) against the ionization energy
% \footnote{\textbf{Here, “ionization energy” denotes the production potential, i.e.\ the minimum photon energy required to ionize the preceding ionic species into the ion of interest (e.g., C~{\sc iii}$\rightarrow$\Civ: 47.9 eV; He~{\sc i}$\rightarrow$He~{\sc ii}: 24.6 eV). So for neutral atoms such as H~{\sc i} and O~{\sc i}, the ionization energy is 0. All values are used consistently throughout this work. These ionization energies can be retrieved from the NIST Atomic Spectra Database at https://physics.nist.gov/asd.}}
of the ions corresponding to each emission line. The Pearson correlation analysis (coefficient $\rho = 0.788$, null-hypothesis probability $p$-value = $0.035$) reveals a clear positive correlation between the attenuation factor and ionization energy, although a moderate degree of scatter is observed in their relationship. It should be noted that \lya\ and \nv\ (represented by blue crosses in Figure~\ref{ionization_energy}) are not included in the correlation analysis because of the blending and absorption issues (see discussions in Section~\ref{sec:measurement}). In addition, the \Civ\ emission line is included in the correlation analysis but with a slightly different method of calculating the attenuation factor. Since we use such a cutoff criterion EW(\Civ)~\textless~$8.9\pm0.2$~\AA\ as the definition of WLQ, the EW(\Civ) distribution of WLQs will certainly not follow a lognormal distribution. We choose to use the $\sigma$ value corresponding to the median EW(\Civ) of WLQs in the lognormal distribution of the full quasar population as the attenuation factor for the \Civ\ emission line (represented by the red cross in Figure~\ref{ionization_energy}). If the \Civ\ emission line is excluded in the correlation analysis, the $\rho$ and $p$-value would become 0.812 and 0.095, respectively. Our conclusion remains unchanged. 

%\lya\ and \nv\ are represented by blue crosses, \civ\ by a blue cross, and other emission lines by red triangles.
%We also calculated their $1 \sigma$ confidence intervals, which are 0.734 and 0.934, 0.002 and 0.060, respectively.

%Namely, the spectra are shifted to the rest-frame, aligned on a common wavelength grid, and normalized at a rest-frame wavelength of 3000  ̊A. They are then stacked by taking the median at each spectral bin.
Figure~\ref{ionization_energy} reveals that for WLQs, generally, the emission line with higher ionization energy has a larger attenuation factor, such as \Civ\ (64.5 eV) and \Heii\ (54.4 eV). On the contrary, the emission line with lower ionization energy has a smaller attenuation factor, typically examples as \Mgii\ (15.0 eV) and \Oi\ (13.6 eV). In other words, the HILs of WLQs are generally much weaker than those of normal quasars, while the LILs are only slightly weaker or comparable. %Moreover, the correlation between WLQ other emission line attenuation factor and corresponding ionization energy is very strong, and the Pearson correlation coefficient reaches 0.788, which can be well explained by shielding gas model.

%The shielding gas model was first proposed in \cite{2011ApJ...736...28W} and \cite{2012ApJ...747...10W} to explain that nearly half of the WLQs are X-ray weak by a factor of $\gtrsim$ 3.3 in the similar optical/UV luminosity and stacking X-ray spectrum of X-ray weak WLQs is hard, a fact further confirmed by \cite{2015ApJ...805..122L} and \cite{2018MNRAS.480.5184N}. 
The different behaviors between HILs and LILs of WLQs, coupled with the positive correlation between line attenuation factor and ionization energy, emerge as natural consequences of the shielding gas scenario. As detailed in Section~\ref{sec:intro}, the  high-energy ionizing photons (mainly in soft X-rays) generated close to the central SMBH are mostly absorbed by the shielding gas; very few of them reach the BLR. Therefore, the number of high-ionization ions in the BLR is greatly reduced. Conversely, the LILs remain relatively unaffected because the outer accretion disk itself can also emit UV photons as the ionizing continuum. 

\begin{figure*}[tp]
	\includegraphics[scale=0.50]{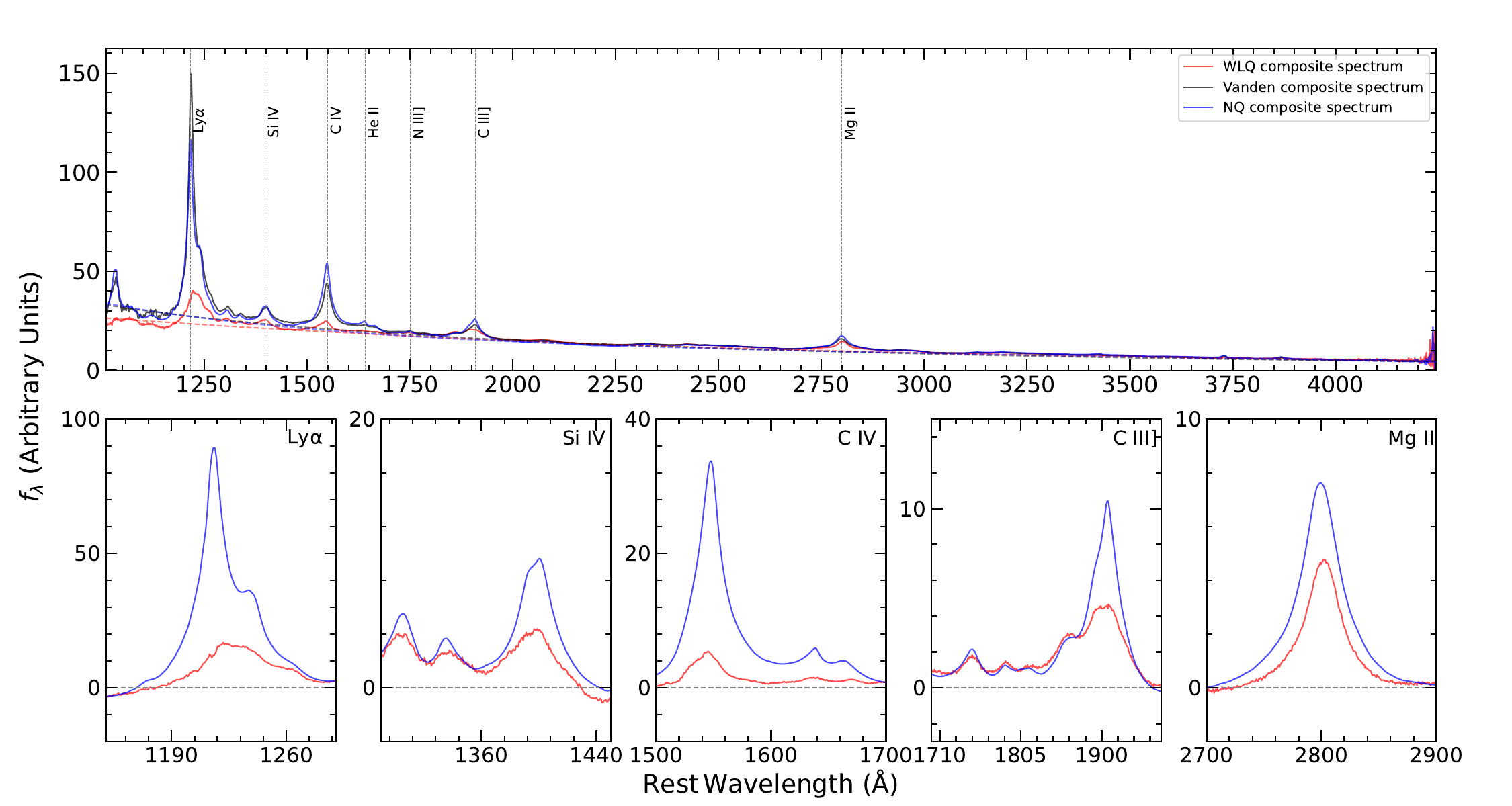}\caption{Comparing the median composite spectra of our WLQ sample, normal quasar sample and that from \citet{2001AJ....122..549V}, all normalized at 2500\AA, which are represented by the red, blue, and black lines, respectively, in the upper panel. Major emission lines are marked with vertical dotted lines. The dashed red, blue, and black lines illustrate the best-fit continua corresponding to each median composite spectrum. In the lower panels, we display five emission lines or complexes for WLQs and normal quasars after continuum subtraction, focusing on \Lya, \siiv, \Civ, \Ciii, and \Mgii, with one panel for each line (complex). The composite spectrum from \cite{2001AJ....122..549V} is excluded from comparison in the lower panels due to its lower redshift and higher luminosity on average, which differ significantly from our WLQ and normal quasar samples.}
	\label{comp_spec_reproduced}
\end{figure*}

To further illustrate the differences between the optical/UV spectra of WLQs and normal quasars, we construct the composite spectra for them, as shown in Figure~\ref{comp_spec_reproduced}. The upper panel presents the median composite spectra for our WLQ sample, normal quasar sample, and that from \cite{2001AJ....122..549V}, represented by the red, blue, and black solid lines, respectively. The spectral slopes $\alpha_{\lambda}$ for these spectra are $-0.85$, $-1.16$ and $-1.08$, respectively.
Major emission lines are labeled by vertical dotted lines, while the dashed red, blue, and black lines represent the best-fit continua for each median composite spectrum. In the lower panels, we compare five emission line complexes in the WLQ and normal quasar composite spectra after subtracting the continua, focusing on \lyanv, \siiv, \Civ, \Ciii\ and \Mgii, with one panel per line complex. In the shielding gas scenario, we can predict that the optical/UV spectrum of WLQs should exhibit a continuum similar to that of normal quasars, while the broad emission lines show varying attenuation factors depending on the ionization energy of the corresponding ion.  
We do not compare the emission lines in the \citet{2001AJ....122..549V} composite spectrum with those of WLQs and normal quasars in the lower panels, as those quasars \citet{2001AJ....122..549V} have substantially lower average redshift and luminosity than those of our WLQ and normal quasar samples, which may result in biased line strength due to the Baldwin effect. 
%in they are not drawn from the same sample. In \cite{2001AJ....122..549V}, the median redshift is 1.253 and major absolute $r'$ magnitude range is between  $-20.0$ to $-25.0$. In contrast, our WLQ and normal quasar samples have a median redshift of 2.201, and the absolute $i$-band magnitude (with $K$ corrections normalized to a redshift of 2) for most sources ranges from $-22.0$ to $-25.0$. Therefore, the our sources are generally more distant and fainter than those in \cite{2001AJ....122..549V}, which is why we do not compare the Vanden Berk composite spectrum directly with the WLQ and normal quasar composite spectra in the lower part of Figure~\ref{comp_spec_reproduced}. 

% We notice that all complex lines except Ly$\alpha$ complex line of Vanden composite spectrum are weaker than that of normal quasar composite spectrum in upper part of the Figure~\ref{comp_spec_reproduced}, which can be well explained by properties of these two samples. Since sources in our subsample are generally darker, their emission lines are naturally stronger under the influence of the Baldwin effect. However, sources in our subsample are farther away, owing to the large scattering cross-section of Ly$\alpha$ photons, neutral hydrogen column densities of $\rm N_{HI} \textgreater 10^{-18} cm^{-2}$ are sufficient for Ly$\alpha$ to enter into the strong absorption regime, which means they encounter more intergalactic medium on their way to the telescope, causing more Ly$\alpha$ absorption.

As a summary, we list the following points of evidences in support of the shielding gas model:
\begin{enumerate}
	\item Nearly half of the WLQs are X-ray weak by a factor of $\gtrsim$ 3.3 compared to typical quasars with similar optical/UV luminosity, while the other half are X-ray normal. This dichotomy can be unified by the shielding gas scenario, in which the puffed inner disk and thick outflows (due to high accretion rates) serve as the shielding gas absorbing X-ray photons. The X-ray weak WLQs are observed with a high inclination angle (``edge-on") where the line of sights go through the shielding gas, while the X-ray normal WLQs are seen at low inclination (``face-on"). 
		%when the quasar's accretion rate reaches to super-Eddington ratio, the inner disk puffs up to form shielding gas, which prevents high energy photons (e.g., EUV and soft X-ray photons) from the more inner disk and corona from reaching the BLR. Under the assumption that the temperature of the corona is normal, when we observe WLQ with a small inclination, we can see an X-ray weak WLQ and an X-ray normal WLQ with a large inclination, thus we can see that nearly half of the WLQs are X-ray weak in the WLQ population. 
	\item X-ray weak WLQs have much harder X-ray spectra than those of typical quasars, supporting the existence of heavy intrinsic absorption by shielding gas. The column density is estimated to be $N_{\rm H}\gtrsim 10^{23}$~cm$^{-2}$  \citep{2015ApJ...805..122L}. On the other hand, X-ray normal WLQs have softer X-ray spectra, which indicates high accretion rates close or exceeding the Eddington limit based on the relationship between the X-ray power-law photon index and Eddington ratio. 
	\item HILs of WLQs are remarkably weak while LILs show normal strength in the optical/UV spectra. The attenuation factor of emission lines have clear dependence upon their corresponding ionization energy. This relationship aligns with the slim disk geometry, in which the high-energy ionizing photons (mainly in X-rays) are generated close to the SMBH and are covered by the puffed inner disk as seen by the BLR. Meanwhile, the low-energy photons are produced in the outer region of the accretion disk without suffering severe absorption.  
\end{enumerate}

\section{Discussion}\label{sec:discuss}
\subsection{Measurements of \lya\ and \nv}\label{sec:measurement}

%However, by construction, these composite spectra only describe the average properties of QSOs, not the intrinsic variations of individual QSOs. Not accounting for these intrinsic properties can bias estimates of the IGM damping wing (Mortlock et al. 2011; Bosman & Becker 2015). Bradley Greig 2017

%Given the BOSS spectral coverage of 3650 to 10400 and the wavelength of the \lya\ emission line, the redshift of sources that include \lya\ in the DR16 catalog is generally greater than 2. This represents a high-redshift population, which is more likely to experience absorption compared to their lower-redshift counterparts. 
% Generally, at these redshift, a series of discrete narrow absorption features which only occur in the \lya\ emission line blueward of a quasar UV spectrum and are caused by the resonant scattering and absorption of \Lya\ photons due to diffuse amounts of neutral hydrogen, are called \Lya\ forest \citep{1998ARA&A..36..267R}. Larger neutral column density absorbers, such as Lyman limit systems and damped DLAs which are ruled out at the data reduction (see Section~\ref{sec:general}), make it more difficult to measure the intrinsic \lya\ emission line. 

For SDSS quasars, the properties of the continua and emission lines in the optical/UV spectra can be reliably obtained using the \mytt{PyQSOFit} package \citep{2018ascl.soft09008G,2019MNRAS.482.3288G,2019ApJS..241...34S}. However, for the \lyanv\ complex, as it is only covered by the SDSS spectra for $z>2$ quasars, the continuum blueward of the \Lya\ emission line and the blue wing of the line itself are inevitably absorbed by the neutral intergalactic medium (IGM), i.e., the \Lya\ forest. The absorption characteristics affect the accurate modeling of the power-law continuum and the \Lya\ line profile, as well as the \Nv\ emission line since it is usually blended with \Lya. Therefore, we believe that absorption at the blueward of \Lya\ makes the \Lya\ and \Nv\ EW distributions statistically unreliable for the large sample of high-redshift quasars in the DR16Q catalog. This is indicated by the unique histogram of the EW(\Nv) distribution which is far unlike that of other emission lines. This is similarly found in previous studies. For example, in \citet{2002ApJ...581..912D}, the Baldwin effect of the \Nv\ emission line clearly deviates from the ionization energy-Baldwin effect slope relationship in their Figure~9. \citet{2001AJ....122..549V} also found that the \nv\ emission deviates from the relationship of line velocity offsets relative to the ionization energy.

Furthermore, emission lines such as \Nv, N~{\sc iv}], and N~{\sc iii}] are particularly sensitive to the nitrogen abundance in AGN spectra (e.g., \citealt{2002ApJ...581..912D}). Since nitrogen is a secondary element, its abundance scales approximately as $\rm N/H \propto (O/H)^2$ (\citealt{2002ApJ...564..592H}). As a result, enhanced nitrogen line strengths can arise even without extreme overall metallicity. \cite{2017A&A...608A..90M} studied a subclass of nitrogen-rich quasars (N-loud QSOs) and found that their strong broad nitrogen lines are primarily driven by an elevated N/O ratio, rather than extreme BLR metallicity. This may naturally account for the observed positive correlation between Eddington ratio and nitrogen abundance (e.g., \citealt{2011A&A...527A.100M}), as N-loud QSOs typically exhibit higher accretion rates. Given that WLQs are often associated with rapid accretion (see discussion in Section \ref{subsec:ledd}), it is plausible that a similar nitrogen enrichment effect is present in this population, potentially influencing the EW distribution of WLQs. This may partially explain why the N~{\sc iii}] and \Nv\ emission lines in WLQs are not as weak as one might expect from pure photoionization arguments.

In light of nitrogen overabundance in high Eddington ratio sources, and considering the severe blending between \Lya\ and \Nv, as well as the fact that blueward absorption in \Lya\ may complicate line decomposition, we choose not to include \Lya\ and \Nv\ in our correlation analysis between emission line attenuation factor and the ionization energy of the corresponding ion. %Furthermore, \Nv\ emission may be selectively enhanced in high-accretion sources due to nitrogen overabundance, making it more difficult to interpret its suppression as purely due to changes in ionization structure.}

% In summary, we believe that when the \lya\ emission line of some sources are not so strong but are absorbed at the same time, the sources with smaller \Lya\ EW will increase, that is, the tail length of the \Lya\ EW distribution will increase. 
% Since PyQSOFit does nothing with the \Lya\ complex of such absorbed sources, the \Lya\ EW must be too small when fitting the spectra of the sources where these blue wings are absorbed, and at the same time indirectly confuses the \nv\ measurement.
%; on the other hand, the absorption characteristics of the blue wing of \Lya\ emission line affect the Gaussian fit of \lya\ and \nv\ emission line. The blue boundary of their continuum fitting window is between 1150 and 1170, which leads to inaccurate continuum fitting when some sources are absorbed in the blueward of Ly$\alpha~\lambda$1216. In addition, they used three Gaussian components of \Lya\ and one component of \Nv\ to fit the \Lya\ complex (its fitting window is 1150-1190\AA) at the same time, which resulted in that the \Lya\ and \Nv\ ~fitting would be affected if the blue wing of \Lya\ was absorbed but without any reconstruction.

A variety of techniques have been proposed to reliably model the intrinsic profile of the \Lya\ emission line and continuum in its blue side for high-redshift quasars. We roughly group these methods into the following categories (see \citealt{2021MNRAS.503.2077B} for a more detailed comparison). The first category of techniques models the continuum and line profiles using the spectrum redward of \Lya\ and extrapolates to the blue side (e.g., \citealt{2006AJ....132..117F, 2015MNRAS.447..499M, 2017MNRAS.466.1814G, 2020ApJ...904...26Y}). The second category is the principal component analysis (PCA) method, which develops a set of eigenvectors to reliably describe the quasar continua, e.g., by constructing a projection matrix \citep{2005ApJ...618..592S, 2011A&A...530A..50P, 2018ApJ...864..143D}, the mean-flux-regulated principal component analysis (MF-PCA) continuum fitting \citep{2012AJ....143...51L}, and the machine learning approach using artificial neural networks \citep{2020MNRAS.493.4256D}. In addition, machine learning approaches have also been developed in recent years (e.g., \citealt{2020MNRAS.493.4256D,2021MNRAS.502.3510L,2023ApJS..269....4S}). If one or more of the methods can be applied to reconstruct the profile of the \lyanv\ complex and its blue-side continuum for SDSS DR16 quasars, we may be able to remodel the \Lya\ and \Nv\ emission line separately and re-calculate their attenuation factor to see whether these two lines follow the correlation between ionization energy and attenuation factor. According to the shielding gas model and nitrogen overabundance in high accretion rate scenario, we would expect that the attenuation factor of \Lya\ is similar to that of \Mgii\ and \Oi, while the strength of the \Nv\ emission line is not as suppressed as predicted purely based on the changes in ionizing photon energies. This is clearly beyond the scope of the current work, and we expect further future studies in this perspective for WLQs. 

\subsection{Do WLQs Represent Quasars with High \ledd ?}\label{subsec:ledd}

In the shielding gas model for WLQs, the central SMBH is accreting close to or exceeding the Eddington limit. The accretion disk can be described with the slim disk model \citep{1988ApJ...332..646A}, in which the radial velocity of the accretion flow becomes comparable with the Keplerian rotation, leading to strong photon trapping and a saturated luminosity. Therefore, the inner disk is puffed up significantly, serving as the shielding gas to prevent high-energy X-ray photons reaching the BLR, resulting in weak HILs \citep{2011ApJ...736...28W, 2015ApJ...805..122L}. Observational evidences supporting this scenario have been listed in Section~\ref{sec:other_emission_lines}. Our methodology of a physically-meaningful definition of WLQs reveals that WLQs and normal quasars may be in different accretion states, with the latter accreting at the standard thin disk regime \citep{1973A&A....24..337S}. The decrease of EW(\Civ) may reflect the transformation of quasar accretion disks from the standard disk to the slim disk state. The bridge quasar sample may be the manifestation of the transition phase (see Figures~\ref{blueshift_weakness(1350)}, \ref{EW_weakness}, and \ref{x_weakness}). 

Based on the above assumptions, we would expect that the Eddington ratios of WLQs should be higher than those of normal quasars. We adopt the Eddington ratio values \ledd\ from the WS22 catalog and directly compare the \ledd\ distributions for WLQs and normal quasars. Given the redshift range of the WLQ and normal quasar samples, the \Hb-based single-epoch SMBH virial mass is not available. Meanwhile, the \Civ-based SMBH virial mass is known to be biased for individual objects (e.g., \citealt{2008ApJ...680..169S}). Therefore, we use the \ledd\ values calculated using the \Mgii-based SMBH virial mass, which is available for quasars in the redshift range of $\sim$0.7--1.9.  The WLQs and normal quasars used in comparison are within the redshift range of 1.45--1.9; their numbers are 1776 and 67635, respectively.

\begin{figure}[tp]
	\includegraphics[scale=0.35]{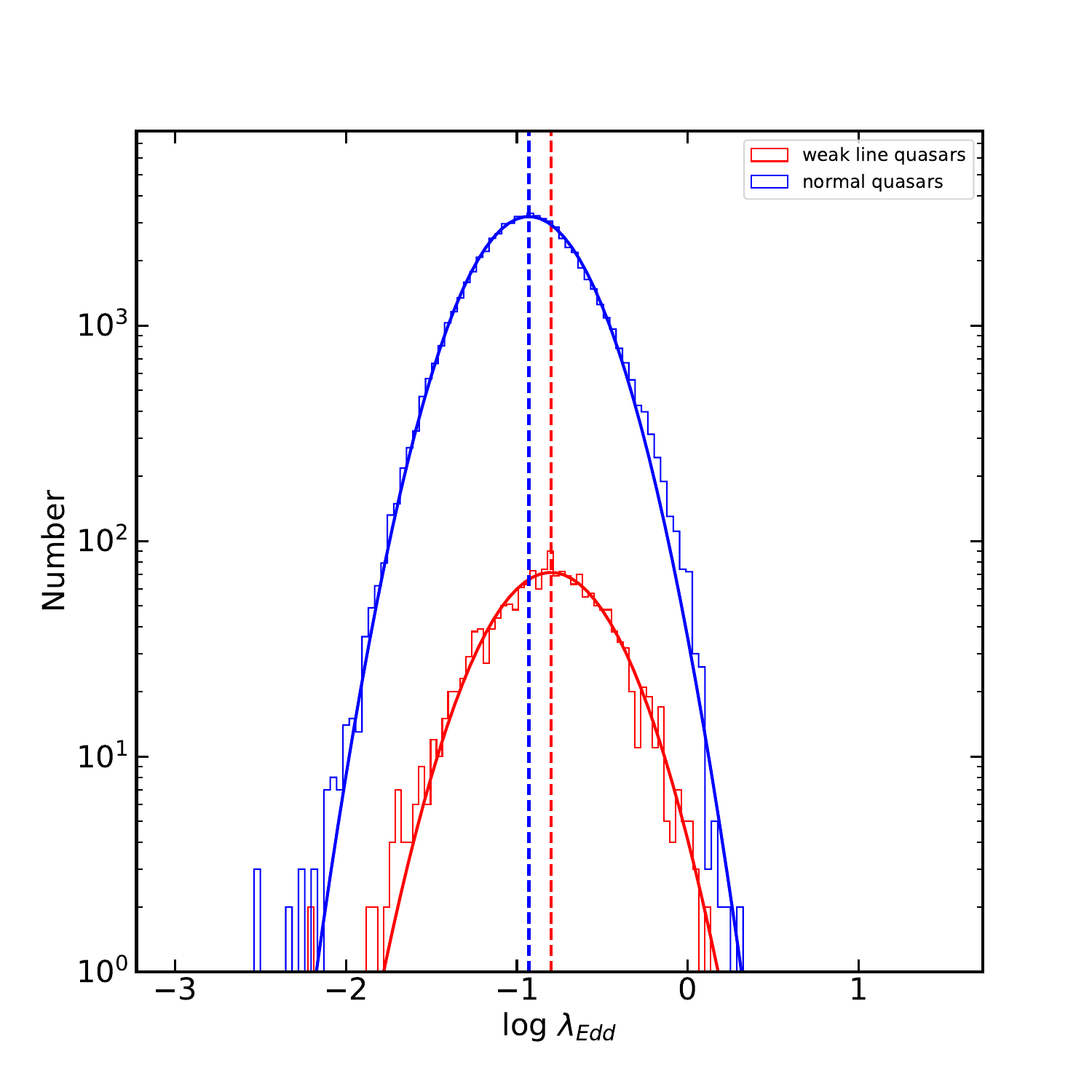}\caption{The Eddington ratio $L/L_{\rm Edd}$ distributions for our WLQ and normal quasar samples, represented by the red and blue histograms. The red and blue lines shows their respective best-fit lognormal models, with the mean values depicted by the vertical dashed lines. %The mean Eddington ratios and 3$\sigma$ range of normal quasar and WLQ in logarithmic coordinates are -0.925 and -0.822, -1.88 $\sim$ 0.03 and -1.87 $\sim$ 0.23, respectively.
    }
	\label{eddington_ratio}
\end{figure}

The \ledd\ distributions for WLQs (red histogram) and normal quasars (blue histogram) are compared in Figure~\ref{eddington_ratio}, with the best-fit lognormal models overlaid. The red and blue dashed lines label the mean $\log$\ledd\ values for WLQs ($-0.822$) and normal quasars ($-0.925$), respectively. The corresponding mean Eddington ratios are \ledd~$=0.151$ for WLQs, and \ledd~$=0.119$ for normal quasars. This result, taken at the face value, appears to indicate that in the $z=1.45$--1.9 range, the overall Eddington ratios of WLQs are not much higher than those of normal quasars as expected. 

This is in contrast with the lines of evidences showing high Eddington ratios for WLQs. 
\citet{2004ApJ...607L.107W} and \citet{2008ApJ...682...81S} show that the hard X-ray flux of quasars is inversely correlated with the \Hb-based \ledd, while the hard X-ray photon index $\Gamma$ is positively correlated with \ledd, suggesting that the release of gravitational potential energy in quasar's hot corona is controlled by the accretion rate. It is well established that over a wide range of redshifts, the radio-quiet type 1 quasars exhibit a typical value of the photon index $\Gamma\sim1.9$ \citep{2005A&A...432...15P, 2005MNRAS.364..195P, 2005AJ....129.2519V, 2007ApJ...665.1004J, 2009ApJS..183...17Y}. \citet{2015ApJ...805..122L} found that the 18 X-ray normal WLQs in their sample have an average $\Gamma$ of $2.18\pm0.09$, while \citet{2018ApJ...865...92M} modeled X-ray spectra for 10 X-ray normal WLQs and found their $\Gamma$ values are generally located in the right 3$\sigma$ tail of photon index distribution of 85 selected radio-quiet type 1 quasars. These results indicate that WLQs may be in a state of high accretion rate from the perspective of photon index as an indicator of \ledd. On the other hand, the NLS1s have similar spectroscopic properties as WLQs in the optical/UV band, i.e., weak broad emission lines (e.g. \civ, \lya\ + \nv) and identical continuum power law slope as that of normal quasars \citep{2023MNRAS.518.6065J}. They generally have a smaller black hole mass of $10^{6}-10^{7} M_{\odot}$, viewed as the low-redshift, low black hole mass counterparts of WLQs. NLS1s are found to be accreting at or above Eddington limit, such as PHL~1811 \citep{2007ApJ...663..103L,2007ApJS..173....1L}, and RX~J0134.2$-$4258 \citep{2022MNRAS.512.5642J, 2023MNRAS.518.6065J}. 

To reconcile this discrepancy, the methodology of SMBH mass measurements should be reconsidered. The single-epoch spectroscopic SMBH mass estimators are based on the assumptions of the virialization of the gas cloud in the BLR and the disk radius-luminosity ($R-L$) relation (e.g., \citealt{2000ApJ...533..631K, 2006ApJ...641..689V, 2013BASI...41...61S}), which may not be applicable to super-Eddington accreting and/or disk-wind dominated quasars. \citet{2014ApJ...797...65W} predicted that super-Eddington quasars should significantly deviate from the normal \Hb\ time lag-luminosity ($\tau_{\rm BLR}-L$) relation, which is caused by the self-shadowing effect, depending on the funnel opening angle and accretion rate. Moreover, \cite{2018ApJ...856....6D} confirmed this prediction and found that for AGNs with super-Eddington accretion, the classical $R–L$ relation can lead to an overestimate of the SMBH mass and thus an underestimate of accretion rate and Eddington ratio. \citet{2022MNRAS.515..491M} found that the overestimation of the SMBH mass can be up to an order of magnitude, which is confirmed by \citet{2023ApJ...950...97H} using 18 WLQs with the \feii-corrected \ledd\ provided. This may explain why the \ledd\ values of WLQs calculated using the \Mgii-based single epoch virial SMBH mass are only slightly higher than those of normal quasars. Moreover, the non-virialization of the BLRs in disk-wind dominated quasars may also contribute to the biased SMBH mass estimates for WLQs \citep{2013ApJ...778...50K}.

Based on the above discussions, we believe that the accretion rate of WLQs is underestimated, and the current Eddington ratio calculation based on the $R-L$ relation and the BLR virialization assumption is not a good indicator for objects such as WLQs which occupy the extreme regions in the $\log$EW(\Civ)-\Civ\ blueshift parameter space. More reliable SMBH measurements can be obtained by applying the correction based on the optical \feii\ emission in 4434--4684~\AA, following \citet{2019ApJ...886...42D} and \citet{2023ApJ...950...97H}. Furthermore, since the radiation efficiency differs in the slim disk and standard disk regimes, the mass accretion rate also serves as a robust indicator to characterize the accretion state of WLQs. We propose to use the dimensionless accretion rate, $\dot{\mathscr{M}}$, calculated by $\dot{\mathscr{M}} = {\dot{M}_{\bullet}}/({L_{\mathrm{Edd}} c^{-2}})$, where $\dot{M}_{\bullet}$ is the mass flow rate, $L_{\mathrm{Edd}}$ is the Eddington luminosity and $c$ is the speed of light. \citet{2016ApJ...818L..14D} developed a method to calculate $\dot{\mathscr{M}}$ for low-redshift quasars, which relates to the ratio of optical \feii\ to \Hb\ flux ($\mathcal{R}_{\rm Fe}$) and the ratio of line FWHM to dispersion for \Hb\ ($\mathcal{D}_{\rm H\beta}$). Obtaining both the reliable Eddington ratio and mass accretion rate would enable us to compare the locations of WLQs and normal quasars in the log$\dot{\mathscr{M}}$-log$L_{\mathrm{Edd}}$ parameter space (see Figure~2 of \citealt{2000PASJ...52..499M}). However, given the redshift range of our WLQ sample ($z>1.45$), near-infrared spectroscopy covering the optical \feii\ and \Hb\ emission are required. We will carry out this investigation in future studies. 

\section{Summary} \label{sec:summary}

In this work, we investigate the emission-line properties of type 1 quasars in the DR16Q catalog, aiming to obtain a physically meaningful definition for WLQs. The EW statistical distributions are presented for a variety of broad and narrow emission lines. We inspect several classical relationships for quasars, including the UV luminosity$-$\Civ\ blueshift relation, the Baldwin effect, and the UV luminosity$-\alpha_{\rm ox}$ relation. We find that the fraction of outliers of these relations depends on the EW values of the \Civ\ emission line, based on which we define the WLQs and normal quasars, as well as the bridge quasars as the transition state of the two. With these definitions, we study the relationship between the emission line attenuation factor of WLQ and the corresponding ionization energy in the optical/UV band. Our main conclusions of this work are summarized as follows:   

%This work is mainly divided into two parts. In the first part (Section~\ref{sec:results}), we use the data in the DR16Q and WS22 catalog to study the dependence between EW(\Civ) and the fraction of outliers in luminosity$-$\Civ\ blueshift, Baldwin effect, and luminosity$-\alpha_{\rm ox}$ relationships, so as to obtain the definition of WLQ and bridge quasars. In the second part (Section~\ref{sec:other_emission_lines}), we use this definition to study the relationship between the emission line attenuation factor of WLQ and corresponding ionization energy in the optical/UV band. Finally, we summarize the main conclusions of this work as follows:

\begin{enumerate}
	\item WLQs are defined as quasars with EW(\Civ) $\textless$ $8.9\pm0.2$~\AA. These quasars represent a special subpopulation of type 1 quasars which have blue continua in the optical/UV band but remarkably weak broad HILs (e.g. \civ\ and \heii), slightly weak or normal LILs (e.g. \oi\ and \mgii). Meanwhile, they often show strong \Civ\ blueshift, weaker \Civ\ emission than predicted by the Baldwin effect, and an exceptionally high fraction (nearly half) of X-ray weak objects. In other words, WLQs have very high fractions of outliers in the classical relations we examined.  
    \item We define quasars with EW(\Civ) $\textgreater$ $19.3\pm0.3$~\AA\ as normal quasars, which have ignorable fractions of outliers in the above-mentioned relations. Quasars with EW(\Civ) less than $19.3$~\AA\ but greater than $8.9$~\AA\ are defined as ``bridge quasars'', representing a transition between WLQs and normal quasars. The outlier fractions of bridge quasars decrease rapidly with increasing EW(\Civ).  
	\item For the statistical distributions of line EW values, all broad emission lines show a tail at their weak-line ends but none at the other side, yet narrow lines exhibit the opposite nature, exhibiting a tail at their strong-line ends, but none at the opposite side.  
	\item We quantify the emission line attenuation factor by comparing the line EW distributions of WLQs and normal quasars. We find that the line attenuation factor is positively correlated with the ionization energy of the corresponding ion. 
    %In other words, , which supports the shielding gas model that block most high-energy photons from the inner disk or corona reaching BLR.
	
\end{enumerate}

Our classification of the WLQs, normal quasars, and bridge quasars, as well as the correlation between the emission line attenuation factor and the ionization energy, demonstrates the validity of the shielding gas model to explain the physical nature of WLQs. In this scenario, normal quasars are accreting at relatively low rates, with the accretion disks following the standard thin disk model. WLQs have much higher accretion rates, close to or exceeding the Eddington limit. The accretion disk can be described by the slim disk model. WLQs represent a distinct physical regime instead of simply the extreme tail of the normal quasar population. The bridge quasars may be in the transition phase between the two accretion-disk regimes. For WLQs, the inner thick disk and the strong outflow serve as the ``shielding gas'' blocking the high-energy photons from reaching the BLR; therefore, the HILs are much more attenuated than the LILs because of the lack of ionizing photons. This explains the positive correlation between the line attenuation factor and ionizing energy. Detailed photoionization computations in the slim disk regime have the potential to directly connect emission line strengths (or flux ratios) to accretion parameters, similar to the work of \citet{2025ApJ...980..134W} under standard thin disk geometry. To more accurately estimate the Eddington ratios \ledd\ of WLQs, the SMBH masses from the single-epoch spectroscopic technique need corrections for quasars with high accretion rates, which would require a future near-infrared spectroscopic campaign on a large sample of WLQs. 

\begin{acknowledgments}

We thank Mouyuan Sun, Qingling Ni, Bin Luo, and Zechang Sun for helpful discussions. This work is supported by the National Key R\&D Program of China under grants 2023YFA1607904, and the National Natural Science Foundation of China under grants 12273029 and 12221003 (X.C. and J.W.). 

Funding for the Sloan Digital Sky Survey IV has been provided by the Alfred P. Sloan Foundation, the U.S. Department of Energy Office of Science, and the Participating Institutions. SDSS acknowledges support and resources from the Center for High-Performance Computing at the University of Utah. The SDSS web site is www.sdss4.org.

\end{acknowledgments}

\facility{Sloan (BOSS spectrographs)}
%\facility{BOSS spectrographs on the 2.5-meter Sloan Telescope at Apache Point Observatory}

\software{Astropy \citep{2013A&A...558A..33A, 2018AJ....156..123A, 2022ApJ...935..167A}, matplotlib \citep{2007CSE.....9...90H}, SciPy \citep{2020NatMe..17..261V}, NumPy \citep{2020Natur.585..357H}, piecewise-regression \citep{2021JOSS...6...3859}, PyQSOFit \citep{2018ascl.soft09008G,2019MNRAS.482.3288G,2019ApJS..241...34S} }

% \section{Appendix}
% \begin{deluxetable}{C{2cm} C{4cm} C{4cm}}[h]
% 	\tabletypesize{\small}
% 	\tablecaption{ionization energy \label{ionization}}
% 	\tablehead{
%         \colhead{ion} & 
% 		\colhead{production potential (eV)} & 
% 		\colhead{destruction potential (eV)} 
% 	}
% 	\startdata
% 	H~{\sc i} &  0  & 13.6\\  
% 	He~{\sc ii} &  24.6 & 54.4\\
%         C~{\sc iii} & 24.4 & 47.9\\
%         C~{\sc iv} &  47.9 & 64.5\\
% 	N~{\sc iii} &  29.6 & 47.4\\
% 	N~{\sc v} & 77.5 & 97.9 \\
% 	O~{\sc i} & 0 & 13.6 \\
% 	Mg~{\sc ii} & 7.6 & 15.0\\
%         Si~{\sc iv} & 33.5 & 45.1
% 	\enddata
% 	\tablecomments{Note that some sources may meet more than one criterion, so the total number of excluded sources does not equal the sum of the number of excluded sources for each criterion.}
% \end{deluxetable}

\end{document}